\documentclass[]{aa}
\usepackage{graphicx}
\usepackage{txfonts}
\newcommand{\epsfig}[1]{\resizebox{\hsize}{!}{\includegraphics{#1}}}
\newcommand{\figcap}[1]{\caption[Another figure]{#1}}
\newcommand{\AJ}[3]{{#1}, AJ, \vol{{#2}}, {#3}.}
\newcommand{\ApJ}[3]{{#1}, ApJ, \vol{{#2}}, {#3}.}
\newcommand{\ApJS}[3]{{#1}, ApJS, \vol{{#2}}, {#3}.}

\newcommand{\AandA}[3]{{#1}, A\&A, \vol{{#2}}, {#3}.}

\newcommand{\MNRAS}[3]{{#1}, MNRAS\rm, \vol{{#2}}, {#3}.}
\newcommand{\MmSAI}[3]{{#1}, Mem. Soc. Astron. Ital.\rm, \vol{{#2}}, {#3}.}

\newcommand{\PASP}[3]{{#1}, PASP\rm, \vol{{#2}}, {#3}.}

\newcommand{\vol}[1]{{\mbox{#1}}}

\newcommand{\Rsolar}{\mbox{$R_{\odot}\,$}}

\newcommand{\FeH}{\mbox{[Fe/H]}\,}

\newcommand{\kms}{\mbox{$\mbox{km\,s}^{-1}$}\,}

\begin{document}
\title{The near-IR Surface Brightness Method applied to\\
 six Cepheids in the young LMC cluster NGC1866\thanks{Based on data
          acquired at the Las Campanas Observatory, Chile,
          the Cerro Tololo Inter American Observatory, Chile,
          and the European Southern Observatory, Chile}
	  \thanks{Tables \ref{tab.JKphot}, \ref{tab.allrv} and
          \ref{tab.gamma} are available in electronic form at
          the CDS via anonymous ftp to {\tt cdsarc.u-strasbg.fr (130.79.128.5)}
          or via {\tt http://cdsweb.u-strasbg.fr/cgi-bin/qcat?J/A+A/XXX/YYY}}
	  }

\author{Jesper~Storm\inst{1}
\and
	Wolfgang~P.~Gieren\inst{2}
\and
	Pascal~Fouqu\'e\inst{3}
\and
	Thomas G. Barnes III\inst{4}
\and
	Mat\'ias G\'omez\inst{2}
}
%

\institute{Astrophysikalisches Institut Potsdam,
        An der Sternwarte 16, D-14482 Potsdam, Germany;
	e-mail: jstorm@aip.de
\and
	Universidad de Concepci\'on, Departamento de F\'{\i}sica,
	Casilla 160-C, Concepci\'on, Chile;
	e-mail: wgieren@coma.cfm.udec.cl, matias@astro-udec.cl
\and
	Observatoire Midi-Pyr\'en\'ees, Laboratoire d'Astrophysique (UMR~5572),
	14, avenue Edouard Belin, F-31400 Toulouse, France;
	e-mail: pfouque@ast.obs-mip.fr
\and
	The University of Texas at Austin, McDonald Observatory,
	1 University Station, C1402, Austin, TX 78712-0259;
	e-mail: tgb@astro.as.utexas.edu
	}
\offprints{J. Storm, e-mail: jstorm@aip.de}
\date{Received 15 March 2005 / Accepted 28 May 2005}

\abstract{
  We present new near-IR light curves for six Cepheids in the young
blue LMC cluster NGC1866 as well as high precision radial velocity
curves for ten Cepheids in NGC1866 and two in NGC2031.
For the six Cepheids in NGC1866 with new $J$ and $K$ light curves we determine
distances and absolute magnitudes by applying the near-IR surface
brightness method. We find that the formal error estimates on the
derived distances are underestimated by about a factor of two. 
We find excellent agreement between the absolute magnitudes for the low
metallicity LMC Cepheids with the Period-Luminosity (P-L)
relation determined by the near-IR surface brightness (ISB) method for 
Galactic Cepheids suggesting that the slope of the P-L relations for
low metallicity and solar metallicity samples could be very similar in
contrast to other recent findings. Still there appears to be significant
disagreement between the observed slopes of the OGLE based apparent P-L
relations in the LMC and the slopes derived from ISB analysis of
Galactic Cepheids, and by inference for Magellanic Cloud Cepheids,
indicating a possible intrinsic problem with the ISB method itself.
Resolving this problem could reaffirm the P-L relation as the prime
distance indicator applicable as well to metallicities significantly
different from the LMC value.
}

\maketitle


\section{Introduction}

  Cepheids still provide one of the most important steps on the extragalactic
distance ladder through the Period-Luminosity (P-L) relation. One
distinguished example is the major effort to determine the value of
the Hubble constant to better than 10\% using
the Hubble Space Telescope as presented by Freedman et al.
(\cite{Freedman01}). Still, the zero-point of the P-L relation remains an
issue of heated debate and Freedman et al. (\cite{Freedman01})
chose to adopt a zero-point based on a best estimate of the distance to
the Large Magellanic Cloud of $18.50\pm0.1$~mag, and to assume a
weak, but still ill-defined, effect on the zero-point as a function
of metallicity. 
For a recent summary of the spread of current LMC distance
estimates see e.g. Gibson (\cite{Gibson00}) and 
Benedict et al. (\cite{Benedict02}).

Fouqu\'e et al. (\cite{FSG03}) and Storm et al. (\cite{Storm04b}) 
(S04 hereinafter) have determined direct distances and
absolute magnitudes to a large sample of Galactic Cepheids
using the near-IR surface brightness (ISB)
method as calibrated by Fouqu\'e and Gieren (\cite{FG97}). Unfortunately
the slopes of the derived Galactic P-L relations differ from those
observed in the LMC by OGLE (Udalski et al. \cite{Udalski99}) in the
optical and by Persson et al. (\cite{Persson04}) in the near-IR. If this
effect is caused by the difference in metallicity of the two Cepheid
populations it will have harmful consequences for the use of the P-L
relation as a distance indicator.

 Here we present new data for a sample of short period LMC
cluster Cepheids and extend the ISB analysis to six Cepheids in the blue
populous LMC cluster NGC1866.  This analysis will allow us to compare
directly the absolute magnitude of LMC Cepheids with those for Galactic
Cepheids using exactly the same procedure for both samples. It will 
also provide a first direct comparison of the slope of the P-L relations
for low metallicity stars in the Magellanic Clouds with the P-L relation
for near solar metallicity Galactic Cepheids by including the SMC
Cepheids studied by S04.

Additionally we
will exploit the fact that the NGC1866 Cepheids can all be considered
to be at the same distance so we can measure directly the dispersion of
the measured distances and thereby the true random error estimate for the
method, something which has previously not been possible.

  NGC1866 is a unique object for many types of studies.  It is young
($\approx 100$~Myr, Brocato et al. \cite{Brocato89}), and it is at
the same time metal-poor when compared to similarly young Galactic
stars. Hill et al.  (\cite{Hill00}) find from high resolution spectroscopy
of three giants a value of $\FeH=-0.5\pm0.1$ confirming the photometric
measurement of $\FeH=-0.46\pm0.18$ by Hilker et al. (\cite{Hilker95}).
It is also very populous and thus provides an excellent laboratory for
studying stellar evolution for intermediate mass stars (e.g. Brocato et
al. (\cite{Brocato03}) and references therein).  In the present context the
most notable feature is the large population of cluster Cepheids. The
possible association of Cepheids with NGC1866 was noticed early
by Shapley and Nail (\cite{SN50}) and photographic light curves were
published by Arp and Thackeray (\cite{AT67}) for seven stars, six of which
are analyzed in the present paper.  Additional Cepheids were identified
by Storm et al.  (\cite{Storm88}), Welch et al. (\cite{Welch91})
(W91 hereinafter),
Welch and Stetson (\cite{WS93}) and most recently, using Hubble Space
Telescope data, by Brocato et al. (\cite{Brocato04}). The Cepheids in
the most crowded parts of the cluster are presently too crowded to be
studied accurately from the ground,
still the list of Cepheids presented by Welch
and Stetson (\cite{WS93}) has more than 20 entries of which most have
determined periods. NGC1866 thus provides us with a unique laboratory to
study the Cepheid instability strip, which has already been exploited
by e.g. W91, Bertelli et al.  (\cite{Bertelli93}), and Brocato et al.
(\cite{Brocato04}).

  C\^ot\'e et al. (\cite{Cote91}) performed the first Baade-Wesselink type
analysis of Cepheids in NGC1866 based on optical data, and the present
analysis will build on their observational data as well as our new
data. W91 provide an excellent historical
summary of the research on the interesting cluster NGC1866 up to the
time of that paper.

  Gieren et al. (\cite{Gieren00b}) presented a first near-IR
surface-brightness analysis for the NGC1866 Cepheid HV12198, and we
present here a re-analysis of this star based on the same
precepts as employed for the other Cepheids in the Milky Way and in the
SMC.

\section{The Sample}

For the present work we have obtained $J$ and $K$ light curves as well
as radial velocity curves for a sample of Cepheids in the young blue
populous clusters NGC1866 and NGC2031.
Finding charts for the NGC~1866 stars can
be found in Storm et al. (\cite{Storm88}) (for NGC1866-V4 see W91
though) and for the NGC~2031 stars in Mateo (\cite{Mateo92}).

Only the data for the NGC1866 Cepheids with Harvard Variable (HV) numbers
proved of sufficient quality and completeness to warrant the application
of the infrared surface brightness method.

\section{The optical photometry}

Optical light curves using CCD detectors have been obtained for a
number of NGC1866 Cepheids by Walker (\cite{Walker87}), W91, 
and most recently by Gieren et al. (\cite{Gieren00a}).
As shown by Gieren et al. (\cite{Gieren00a}) these data sets are all in
very good agreement, especially in the $V$-band, and taken together they
provide excellent phase coverage and high accuracy for the six Cepheids
which are of main interest to the present paper. Consequently we have
adopted this combined $BVRI$ data set for further analysis.
To ensure that all the data are on the same system we have applied small
offsets (few hundredth of a mag) to the various data sets to bring them into
optimal agreement with the Gieren et al. data set. 

  The ephemerides which we have used are from Gieren et al.
(\cite{Gieren00a}) for the stars with HV numbers, for the remaining
stars we have determined periods and epochs on the basis of the
available radial velocity data (see also Sec.\ref{subsec.rv}) 
and they are summarized in Tab.\ref{tab.ephem}.
W91 found evidence for a slightly variable period for HV12198 but we do
not find any significant variation over the 11 year period spanned by the
optical data from Walker (\cite{Walker87}), W91, and 
Gieren et al. (\cite{Gieren00a}), so we have simply
adopted a constant period for this star.

\section{The near-IR data}

  We have obtained near-IR photometry in the $J$ and $K'$ bands at ESO, La
Silla, using the ESO-MPG 2.2m telescope with the IRAC-2 camera during
observing runs in Dec.~1996 and Jan.~1998. We have also acquired
photometry in the $J$ and $K$s bands at the Las Campanas Observatory
(LCO) using the 1m Swope telescope with the IRC in Nov.~1998 and 
Jan.~1999, and the 2.5m du~Pont telescope in Jan.~1999.  Finally we used
the OSIRIS instrument at the 1.5m telescope at the Cerro Tololo
Inter-American Observatory in Dec.~2000 to obtain $J$ and $K$s
photometry.

\subsection{Data reduction}

  For all the cameras we applied the appropriate non-linearity
correction as quoted in the manuals. The exposure times were
chosen to avoid exposing the detector to more than half of
full well, so in all cases the applied non-linearity correction remained small.

  The data were obtained in sets where one set typically consisted of
five or six different pointings on the target. In this way a local sky
frame could be determined from the other frames in a set by median
filtering. These local skyframes were subtracted from each of the
original frames. In the case of more crowded fields, i.e. close to the
cluster center, it was necessary to use sky frames based on data sets
from another target with same exposure parameters obtained close in time.

  In case of the La Silla data the flat fields were determined by
subtracting a domeflat with the quartz lamp off from a dome flat with
the quartz lamp on. A number of these exposure sets were averaged
together to achieve a very high signal to noise ratio. A set of
observations of a standard star placed in a grid of 4 by 4 positions on
the chip was used to determine the illumination correction to be applied
to the flatfield exposures and this corrected flatfield was used to
flat field the science exposures.

  In the case of the Las Campanas data we used twilight flatfields. Here
the signal to noise ratio was somewhat lower, as the exposure level
could not be as finely adjusted as was the case for the La Silla data.
No additional correction for illumination effects was included.

  Finally the bad pixels were eliminated using the IRAF {\tt fixpix} routine.

\subsection{Photometry}

  The PSF-fitting photometry was carried out with DoPHOT2.0 
(Schechter et al. \cite{Schecter93}) running on a DEC-Alpha with 
OSF/1 UNIX. 

\begin{figure}[htp]
\epsfig{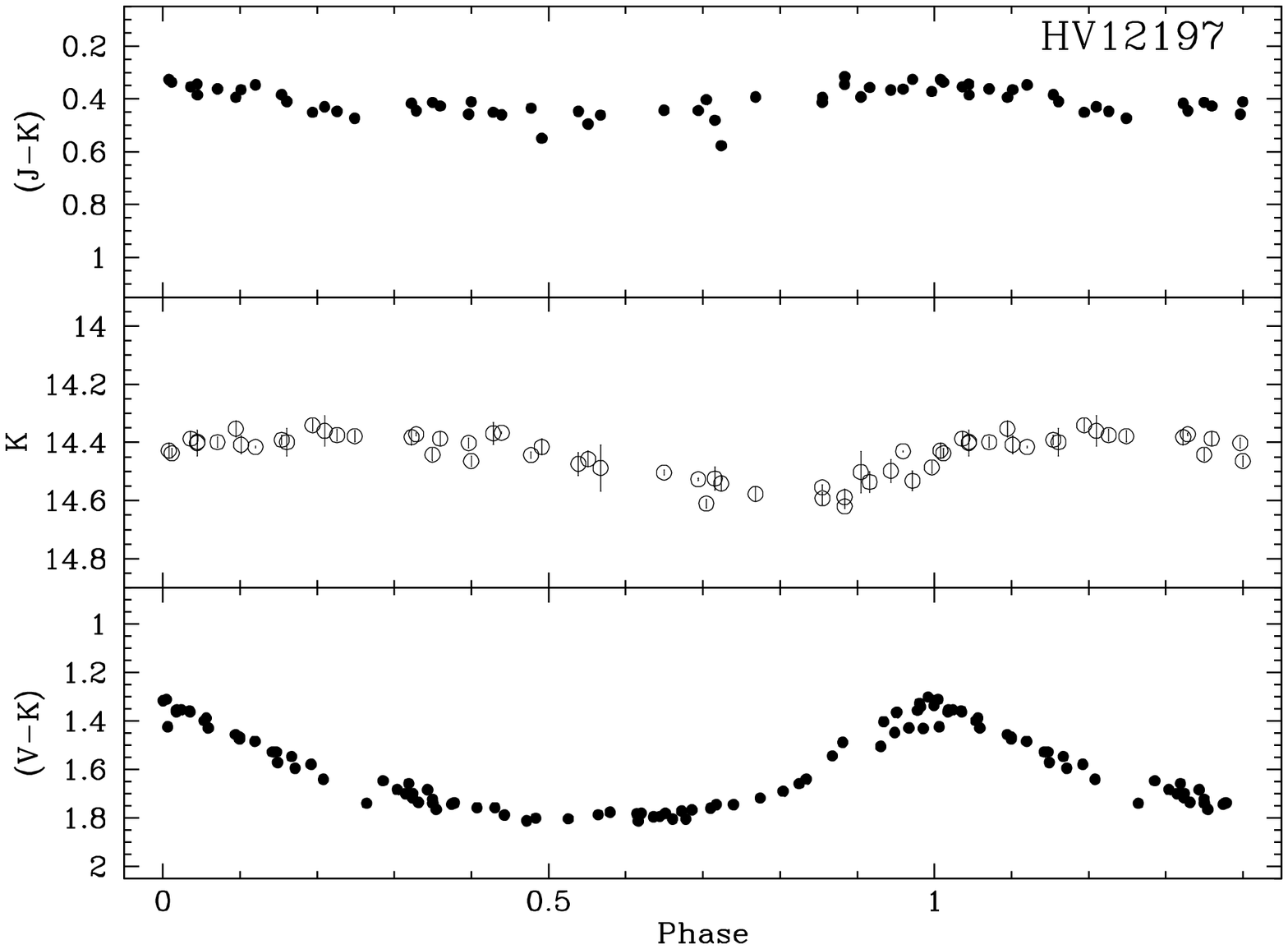}
\figcap{\label{fig.hv12197vvks} K-band light curve and $(V-K)$ and
$(J-K)$ colour curves for HV12197.}
\end{figure}

\begin{figure}[htp]
\epsfig{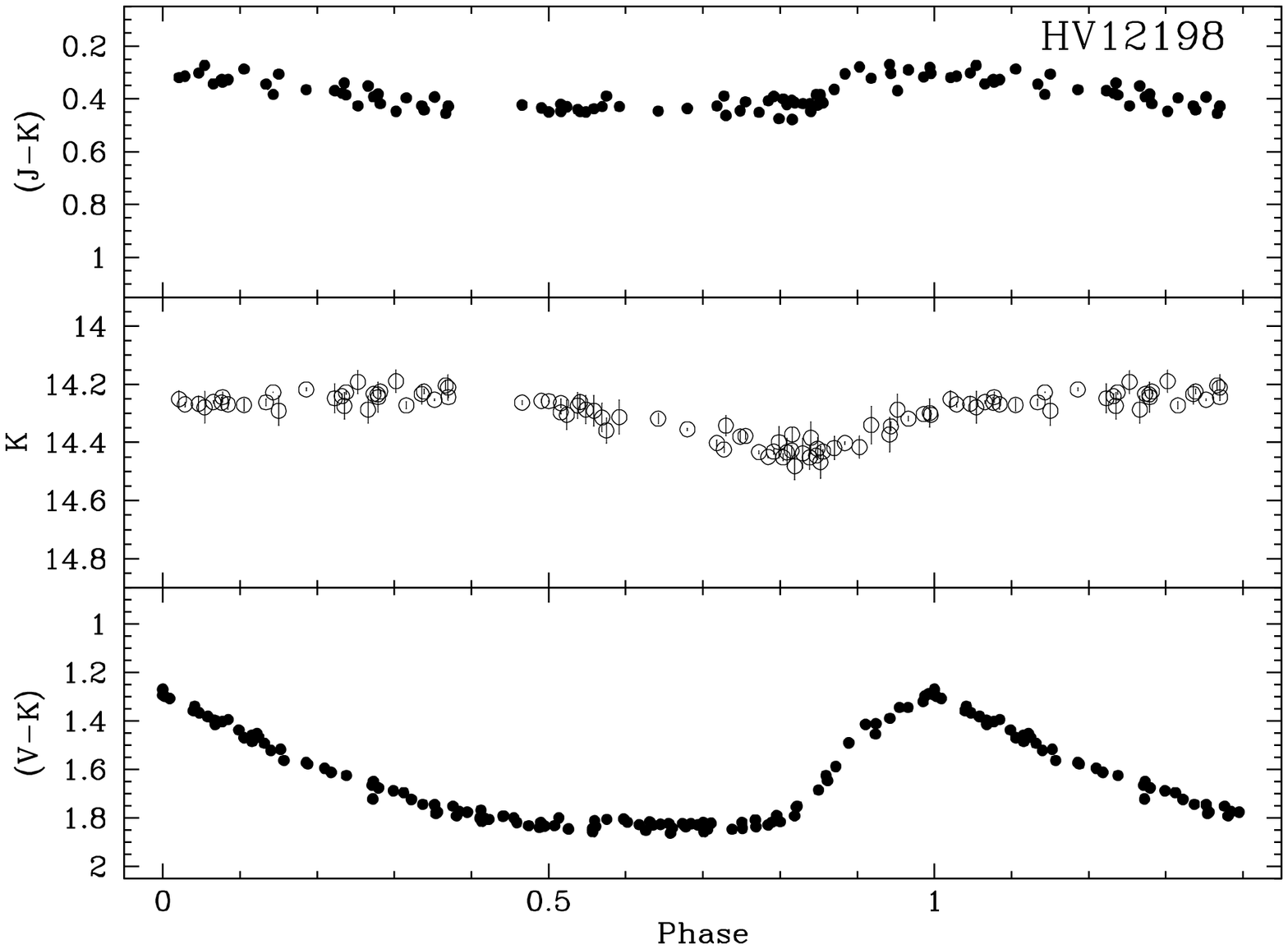}
\figcap{\label{fig.hv12198vvks} K-band light curve and $(V-K)$ and
$(J-K)$ colour curves for HV12198.}
\end{figure}

\begin{figure}[htp]
\epsfig{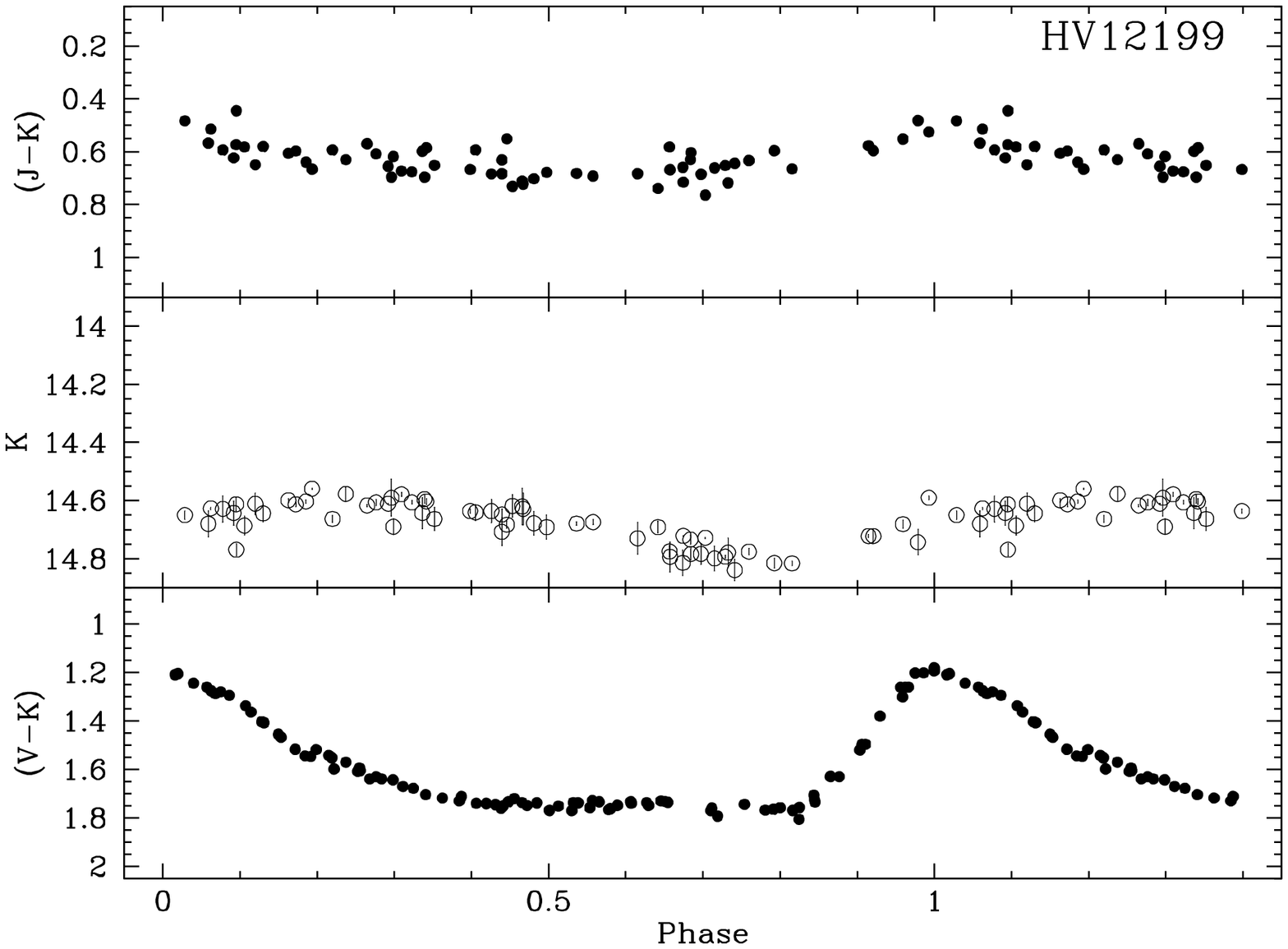}
\figcap{\label{fig.hv12199vvks} K-band light curve and $(V-K)$ and
$(J-K)$ colour curves for HV12199.}
\end{figure}

\begin{figure}[htp]
\epsfig{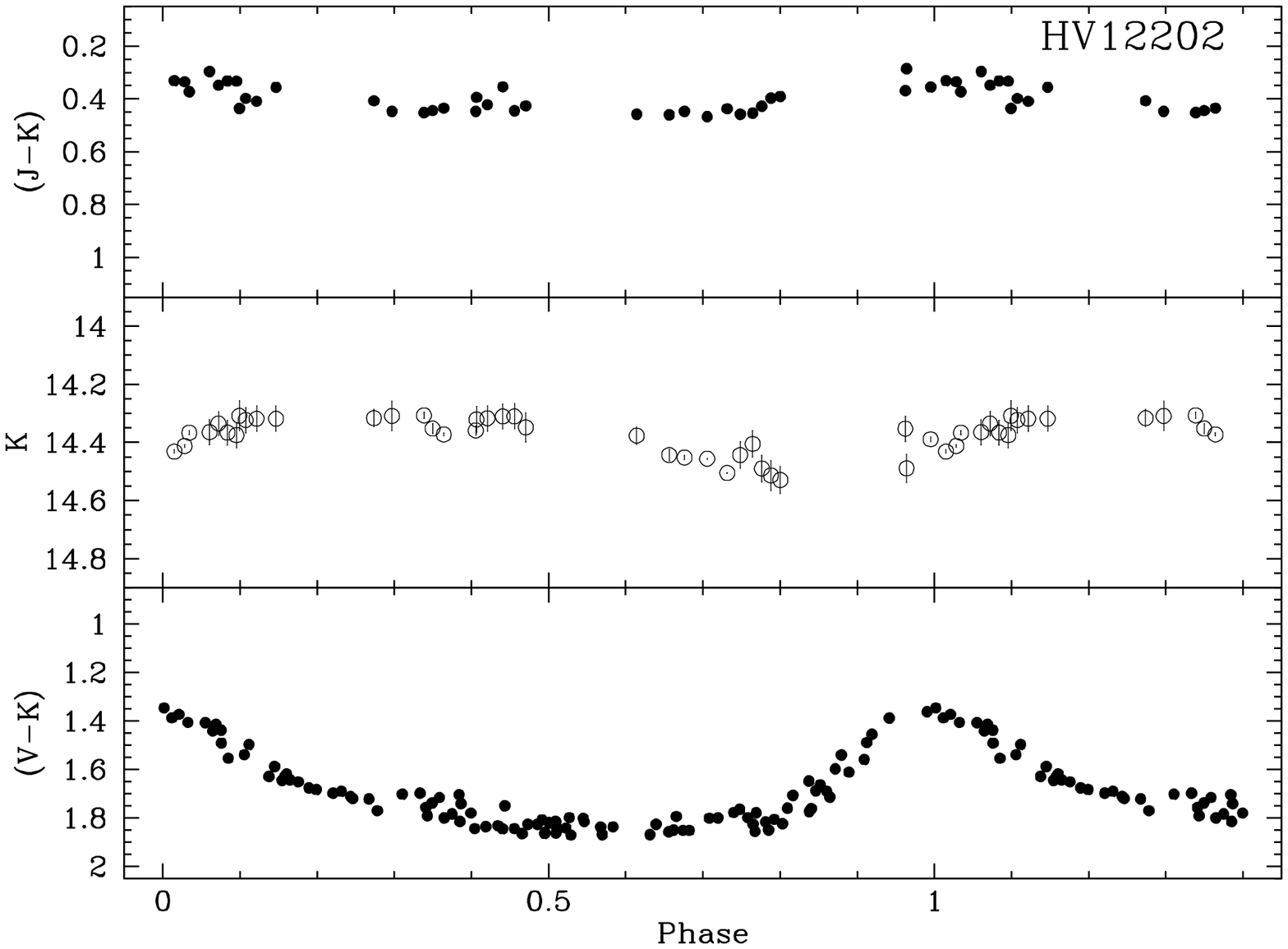}
\figcap{\label{fig.hv12202vvks} K-band light curve and $(V-K)$ and
$(J-K)$ colour curves for HV12202.}
\end{figure}

\begin{figure}[htp]
\epsfig{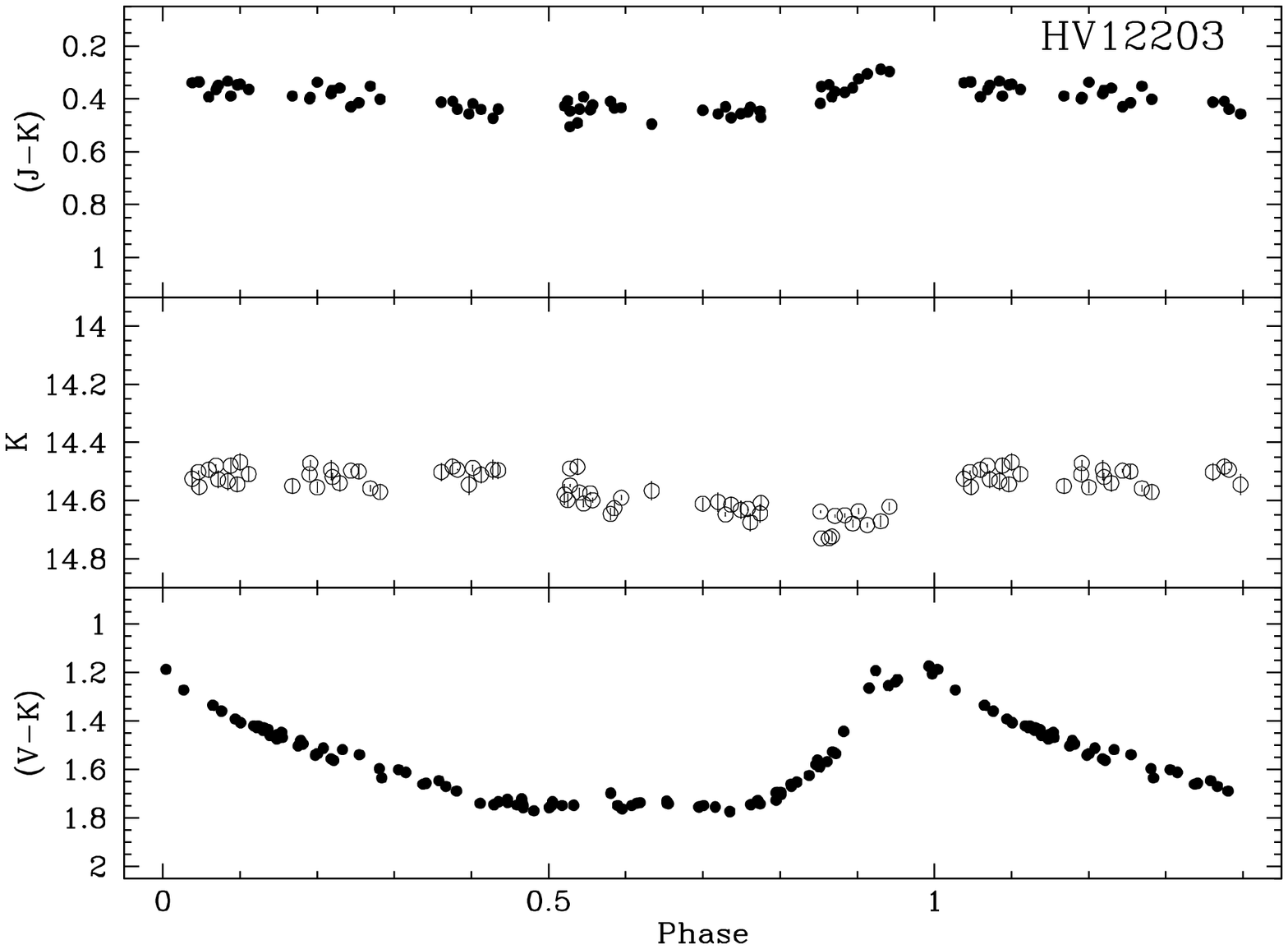}
\figcap{\label{fig.hv12203vvks} K-band light curve and $(V-K)$ and
$(J-K)$ colour curves for HV12203.}
\end{figure}

\begin{figure}[htp]
\epsfig{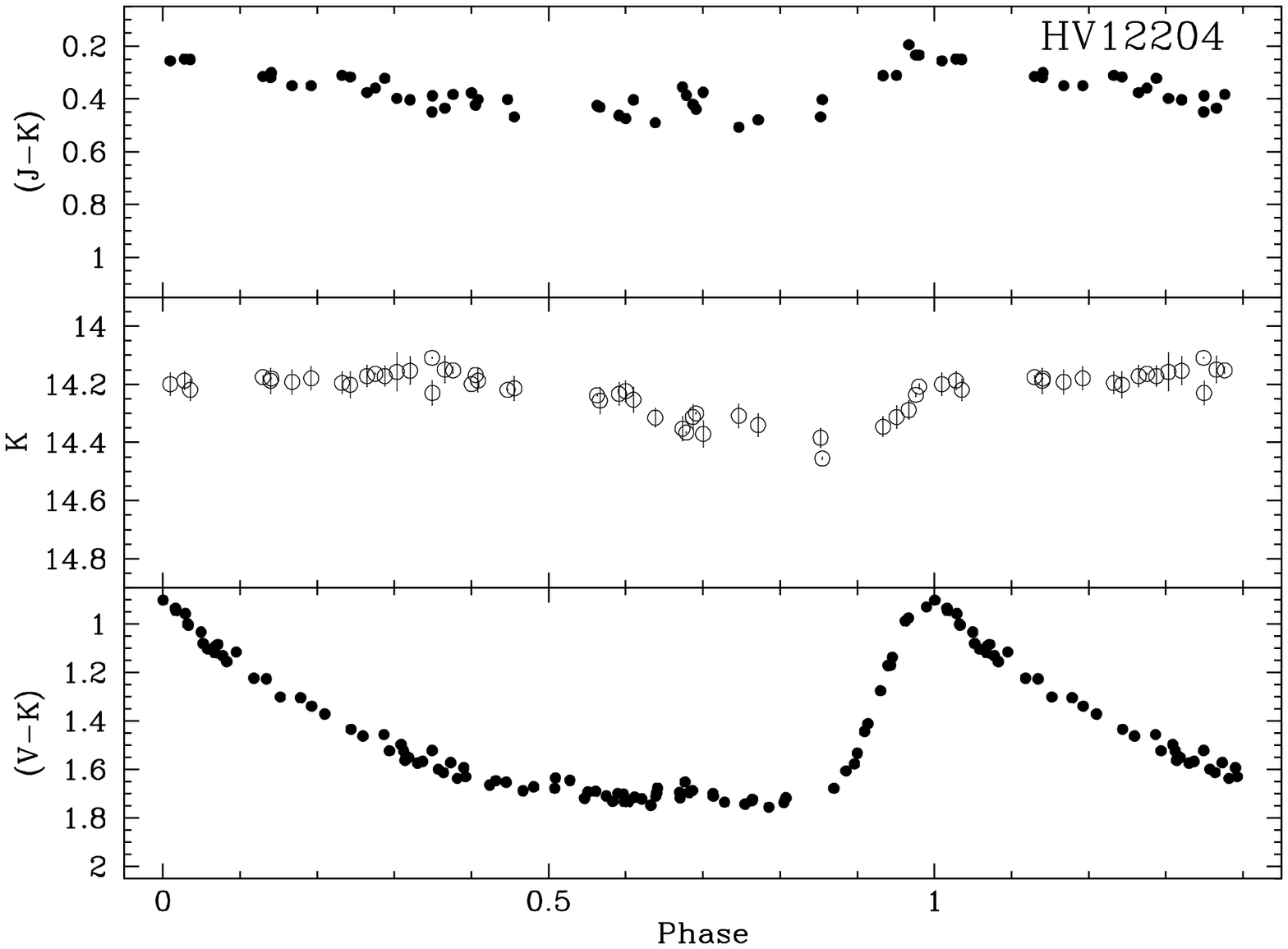}
\figcap{\label{fig.hv12204vvks} K-band light curve and $(V-K)$ and
$(J-K)$ colour curves for HV12204.}
\end{figure}

For each field a database was constructed containing the photometry of
the stars from all the individual exposures of the field. On the basis of
this database comparison stars could be selected and used as a reference
to transform the photometry of each exposure to the instrument system.
The detailed transformation procedure is described in the Appendix and
the calibrated photometric data is tabulated in Tab.\ref{tab.JKphot} and
the resulting light and color curves have been plotted in
Figs.\ref{fig.hv12197vvks}-\ref{fig.hv12204vvks}.

\begin{table}
\caption{\label{tab.JKphot} The $K$ and $(J-K)$ photometric measurements
for the program stars transformed to the CIT system as described in the
text. The complete table is available electronically from the CDS.}
\begin{tabular}{c c c c c c}
\hline\hline
ID & HJD & $K$ & $\sigma_{\mbox{\scriptsize (K)}}$ &
$(J-K)$ & $\sigma_{\mbox{\scriptsize (J-K)}}$ \\
 & [days] & [mag] & [mag] & [mag] & [mag] \\
\hline
HV12197 & 2450438.65133 & 14.524 & 0.039 & 0.481 & 0.045\\
HV12197 & 2450439.57052 & 14.428 & 0.026 & 0.327 & 0.028\\
HV12197 & 2450439.65912 & 14.387 & 0.024 & 0.354 & 0.026\\
HV12197 & 2450439.76887 & 14.399 & 0.018 & 0.362 & 0.020\\
HV12197 & 2450439.86449 & 14.408 & 0.030 & 0.365 & 0.032\\
HV12197 & 2450440.55941 & 14.382 & 0.027 & 0.417 & 0.030\\
HV12197 & 2450440.67597 & 14.387 & 0.022 & 0.427 & 0.027\\
HV12197 & 2450440.79185 & 14.402 & 0.017 & 0.458 & 0.028\\
HV12197 & 2450441.76019 & 14.610 & 0.015 & 0.403 & 0.032\\
cont. & & & & & \\
\hline\hline
\end{tabular}
\end{table}

\section{The radial velocities}

  The radial velocity curves are based on high resolution spectra
obtained with the Las Campanas Observatory 2.5m du Pont telescope
equipped with the echelle spectrograph and the 2D-Frutti photon counting
system. This system provides a dispersion of 0.2nm/mm ($\approx
0.006\mbox{nm~pix}^{-1}$) and was used with a slit of $1.5 \times 4$~arcsec
which transforms into an instrumental profile with a
FWHM of approximately 3.5~pixels or 12~\kms corresponding to a
resolving power of $R\approx 20000$.

We also used the Cerro Tololo 4m Blanco telescope with the echelle
spectrograph, the red long camera and the Tek2048 CCD detector.
This system provides a very similar resolving power ($R\approx 20000$)
with a dispersion
of about 0.25nm/mm which combined with 0.024~mm pixels again gives
about $0.006\mbox{nm~pix}^{-1}$. The slit was again set at 1.5 arcsec
width. Integration times for our science objects with both systems were
between 20 and 45 minutes depending on the weather conditions and the
brightness of the star.

\subsection{The observations}

  The Las Campanas data were obtained during three observing runs in
Dec. 1995, Jan. 1996 and Jan. 2000, and the CTIO data were obtained
during a single run in Jan. 2000. The corresponding Heliocentric Julian
Dates are tabulated in Tab.\ref{tab.allrv} with the derived velocities. 

  The science spectra were interlaced with Th-Ar lamp spectra, and as the
science objects were all in the same part of the sky, most science
exposures had bracketing calibration spectra. A number of radial
velocity standard stars were observed during each run to check the zero
point. During the CTIO run, twilight sky exposures were obtained for the
same purpose. The bright star AT67A (Arp and Thackeray, \cite{AT67}),
earlier selected by W91 to be similar to the Cepheids in spectral type
and with sharp lines, which is located close to the two stellar clusters
was observed frequently and served as our cross-correlation template star.

\subsection{Data reduction}

  The Las Campanas data were first divided by a lamp flat field
to take out the pixel to pixel variation in the detector. The apertures
corresponding to each echelle order were extracted using the
IRAF\footnote{IRAF is the Image Reduction and Analysis
Facility, made available to the astronomical community by the National
Optical Astronomy Observatories, which are operated by AURA, Inc., under
cooperative agreement with the National Science Foundation.}
{\tt noao.imred.echelle.apall} command
where the individual apertures had first been traced on a well exposed
stack of standard star frames. The calibration Th-Ar apertures were
extracted for each science exposure using the apertures previously
defined by the corresponding science exposure.

  The wavelength solution was determined for each calibration
Th-Ar spectrum and using the IRAF {\tt refspectra} and {\tt dispcor}
commands the science spectra were wavelength calibrated by weighting the
bracketing Th-Ar spectra according to their closeness in time to the
science exposure.

  Ten echelle orders covering the wavelength range from
$444-513$~nm were extracted and continuum corrected before they were 
merged into a one-dimensional spectrum using the {\tt scombine} command.
Finally outlying points, mostly from the low signal-to-noise
ends of the spectral orders were removed to eliminate spurious signals.

  The CTIO CCD data were bias subtracted with the bias level
corresponding to each individual quadrant of the quad-read CCD and each
quadrant had its gain normalized. No pixel to pixel variation correction
(flatfielding) was deemed necessary, and we preferred to refrain from such
a correction to avoid the introduction of additional noise into the data.
For the longer exposures it was necessary to remove the cosmic ray hits,
and this was achieved by simply eliminating counts higher than a certain
threshold together with the pixels within a 2 pixel radius.

  The spectral retrieval proceeded as for the Las Campanas data and 10
echelle orders covering the range $480-529$~nm were finally extracted.
The length of the extracted one-dimensional spectra were in both cases
limited by the fact that the cross-correlation procedure used can only
handle spectra with a maximum of 8192 pixels.

  The spectra were cross-correlated numerically using the technique 
developed by Tonry and Davis (\cite{Tonry79}) implemented in the
{\tt rvsao} package (Kurtz and Mink \cite{rvsao98}) within IRAF. 

  For each run the spectra of the local reference star AT67A after
dispersion correction were stacked to provide a high signal to noise
reference template spectrum. The science spectra were then
cross-correlated against this template, for which we adopted a velocity
of $-0.7$~\kms as already found by W91 and
confirmed by our measurements with respect to a twilight sky template
and the standard stars observed.

  Nightly radial velocity offsets were determined on the basis of the shifts
of the individual AT67A observations and/or twilight sky spectra against
the template and these offsets were finally added to the derived
velocities. Using the {\tt rvsao} package the observed radial velocities 
were transformed into barycentric velocities, which are the
values tabulated in Tab.\ref{tab.allrv}. The phases listed in the table
are based on the ephemerides from Tab.\ref{tab.ephem}.

\begin{table*}
\caption{\label{tab.allrv}The barycentric radial velocity measurements,
$V_r$, for each star,
with the associated Heliocentric Julian Date, HJD, and phase, $\phi$. 
The estimated random error for all the velocities is of the order 1\kms.
The complete table is available from the CDS.}
\begin{tabular}{c c c c c c c c c c c c}
\hline\hline
HJD & $\phi$ & $V_r$ &
HJD & $\phi$ & $V_r$ &
HJD & $\phi$ & $V_r$ &
HJD & $\phi$ & $V_r$ \\
$-2400000$ & & &
$-2400000$ & & &
$-2400000$ & & &
$-2400000$ & & \\

[days] & & [\kms] &
[days] & & [\kms] &
[days] & & [\kms] &
[days] & & [\kms] \\

\hline
\multicolumn{3}{c}{\bf HV12197} & 50097.6485 & 0.70 & 320.1 & \multicolumn{3}{c}{\bf HV12204} & 50096.5568 & 0.58 & 304.3 \\
50066.6647 & 0.39 & 298.5 & 51555.7082 & 0.17 & 286.3 & 50066.5654 & 0.15 & 282.2 & 50096.7672 & 0.68 & 306.5 \\
50067.5543 & 0.67 & 314.2 & 51556.5402 & 0.48 & 308.8 & 50066.7376 & 0.20 & 284.1 & 50097.5561 & 0.07 & 293.2 \\
50067.8071 & 0.76 & 315.5 & 51556.6603 & 0.53 & 310.1 & 50067.6675 & 0.47 & 302.3 & 50097.7846 & 0.19 & 289.7 \\
50068.6029 & 0.01 & 283.8 & 51556.8110 & 0.58 & 314.4 & 50068.7943 & 0.80 & 318.9 &  & &  \\
50095.7638 & 0.65 & 310.9 & 51557.6894 & 0.92 & 292.6 & 50096.7144 & 0.92 & 304.9 & \multicolumn{3}{c}{\bf NGC2031-V4} \\
50096.6082 & 0.92 & 290.1 & 51558.7151 & 0.31 & 297.4 & 50097.7381 & 0.21 & 286.8 & 51555.5790 & 0.01 & 224.5 \\
50096.8013 & 0.98 & 281.6 & 51572.7110 & 0.61 & 315.3 & 51555.6398 & 0.18 & 284.3 & 51555.7933 & 0.07 & 228.3 \\
50097.6318 & 0.24 & 292.6 & 51574.7432 & 0.38 & 300.0 & 51556.6832 & 0.48 & 304.0 & 51556.5662 & 0.29 & 243.2 \\
50097.8436 & 0.31 & 292.1 &  & &  & 51558.5679 & 0.03 & 275.1 & 51556.7349 & 0.34 & 245.8 \\
51555.6825 & 0.03 & 282.5 & \multicolumn{3}{c}{\bf HV12202} & 51558.6887 & 0.06 & 276.0 & 51557.7196 & 0.63 & 260.9 \\
51555.8249 & 0.07 & 278.6 & 50066.6821 & 0.43 & 314.6 & 51571.7364 & 0.86 & 323.3 & 51558.5926 & 0.89 & 248.2 \\
51556.7124 & 0.35 & 298.5 & 50067.6265 & 0.73 & 326.5 & 51573.7296 & 0.44 & 302.0 & 51571.6981 & 0.71 & 267.4 \\
51556.8360 & 0.39 & 301.0 & 50067.8286 & 0.80 & 325.7 &  & &  & 51571.8097 & 0.74 & 267.5 \\
51558.5401 & 0.94 & 287.3 & 50068.7466 & 0.09 & 292.0 & \multicolumn{3}{c}{\bf NGC1866-V4} & 51572.5492 & 0.96 & 228.1 \\
51571.5869 & 0.09 & 282.2 & 50095.8325 & 0.83 & 321.1 & 50067.7223 & 0.77 & 310.5 & 51572.6442 & 0.98 & 227.7 \\
51572.7538 & 0.46 & 304.1 & 50096.6756 & 0.10 & 292.1 & 50068.6640 & 0.05 & 293.5 & 51572.7860 & 0.03 & 225.8 \\
51573.5842 & 0.72 & 316.6 & 50097.6901 & 0.43 & 313.2 & 50097.5985 & 0.77 & 312.6 & 51573.5461 & 0.25 & 241.7 \\
51573.7776 & 0.78 & 315.2 & 51571.6203 & 0.72 & 308.1 & 50097.8082 & 0.83 & 306.7 & 51573.6566 & 0.28 & 244.1 \\
 & &  & 51572.6177 & 0.04 & 275.1 &  & &  & 51574.5423 & 0.54 & 259.8 \\
\multicolumn{3}{c}{\bf HV12198} & 51573.6939 & 0.38 & 293.0 & \multicolumn{3}{c}{\bf NGC1866-V6} & 51574.6699 & 0.57 & 262.2 \\
50066.6203 & 0.16 & 284.0 & 51574.6297 & 0.69 & 307.4 & 50067.7598 & 0.72 & 306.2 & 51574.7803 & 0.61 & 263.2 \\
50066.7786 & 0.20 & 289.4 &  & &  & 50068.6848 & 0.20 & 291.2 &  & &  \\
50067.6057 & 0.44 & 304.1 & \multicolumn{3}{c}{\bf HV12203} & 50095.7013 & 0.15 & 293.2 & \multicolumn{3}{c}{\bf NGC2031-V11} \\
50068.7307 & 0.76 & 318.7 & 50066.5904 & 0.27 & 294.1 & 50096.5415 & 0.59 & 309.6 & 51555.5537 & 0.25 & 236.7 \\
50095.8158 & 0.44 & 307.2 & 50066.7585 & 0.32 & 302.1 & 50096.7436 & 0.69 & 311.4 & 51555.7669 & 0.32 & 243.3 \\
50096.6464 & 0.68 & 316.2 & 50067.6466 & 0.62 & 315.4 & 50097.5401 & 0.10 & 290.2 & 51556.5884 & 0.61 & 263.9 \\
50096.8256 & 0.73 & 314.9 & 50067.8480 & 0.69 & 319.6 & 50097.7700 & 0.22 & 295.0 & 51556.7571 & 0.67 & 265.4 \\
50097.6714 & 0.97 & 281.8 & 50067.8637 & 0.70 & 319.2 &  & &  & 51558.6141 & 0.33 & 245.0 \\
51571.5604 & 0.36 & 301.0 & 50068.7692 & 0.00 & 284.8 & \multicolumn{3}{c}{\bf NGC1866-V7} & 51571.6610 & 0.95 & 221.2 \\
 & &  & 50095.8520 & 0.17 & 290.5 & 50066.7189 & 0.65 & 309.6 & 51571.7738 & 0.99 & 213.8 \\
\multicolumn{3}{c}{\bf HV12199} & 50096.6908 & 0.45 & 308.3 & 50067.6876 & 0.94 & 294.1 & 51572.5861 & 0.28 & 241.3 \\
50066.6376 & 0.95 & 286.7 & 50097.7193 & 0.80 & 318.3 & 50068.6265 & 0.21 & 295.4 & 51572.8168 & 0.36 & 244.6 \\
50066.8064 & 0.01 & 278.8 & 51555.6186 & 0.32 & 300.9 & 50096.5853 & 0.47 & 298.2 & 51573.6204 & 0.64 & 263.1 \\
50067.5748 & 0.30 & 293.4 & 51555.7429 & 0.36 & 303.3 & 50096.7818 & 0.52 & 297.9 & 51574.5781 & 0.98 & 220.5 \\
50067.7814 & 0.38 & 303.5 & 51556.6311 & 0.66 & 319.4 & 50097.5797 & 0.76 & 310.8 & 51574.7004 & 0.03 & 222.1 \\
50068.5866 & 0.69 & 319.7 & 51556.7839 & 0.71 & 322.0 & 50097.8276 & 0.83 & 312.2 & 51574.8164 & 0.07 & 223.5 \\
50068.8237 & 0.78 & 316.9 & 51557.6672 & 0.01 & 283.9 &  & &  &  \\
50095.7790 & 0.99 & 277.6 & 51558.6506 & 0.34 & 301.5 & \multicolumn{3}{c}{\bf NGC1866-V8} &  \\
50096.6235 & 0.31 & 295.9 & 51572.6850 & 0.10 & 285.6 & 50067.7022 & 0.20 & 286.6 &  \\
50096.8422 & 0.39 & 299.4 &  & &  & 50068.6428 & 0.67 & 311.7 &  \\
\hline
\end{tabular}
\end{table*}

\begin{table}
\caption{\label{tab.ephem}The ephemerides for the stars.}
\begin{tabular}{r l l}
\hline\hline
ID & Period & Epoch of max. $V$ light\\
& [days] & [HJD] \\
\hline
\object{HV12197} & 3.14381 & 2450071.72 \\
\object{HV12198} & 3.52279 & 2450069.59 \\
\object{HV12199} & 2.63918 & 2450103.72 \\
\object{HV12202} & 3.10112 & 2450071.56 \\
\object{HV12203} & 2.95411 & 2450071.715 \\
\object{HV12204} & 3.43876 & 2450069.49 \\
\object{NGC1866-V4} & 3.31895 & 2450068.5 \\
\object{NGC1866-V6} & 1.93386 & 2450068.3 \\
\object{NGC1866-V7} & 3.388369 & 2450067.9 \\
\object{NGC1866-V8} & 2.00714 & 2450067.3 \\
\object{NGC2031-V4} & 3.42871 & 2451572.7 \\
\object{NGC2031-V11} & 2.8244 & 2451571.8 \\
\hline
\end{tabular}
\end{table}

\subsection{Radial velocity curves}
\label{subsec.rv}

\begin{figure*}[htp]
\epsfig{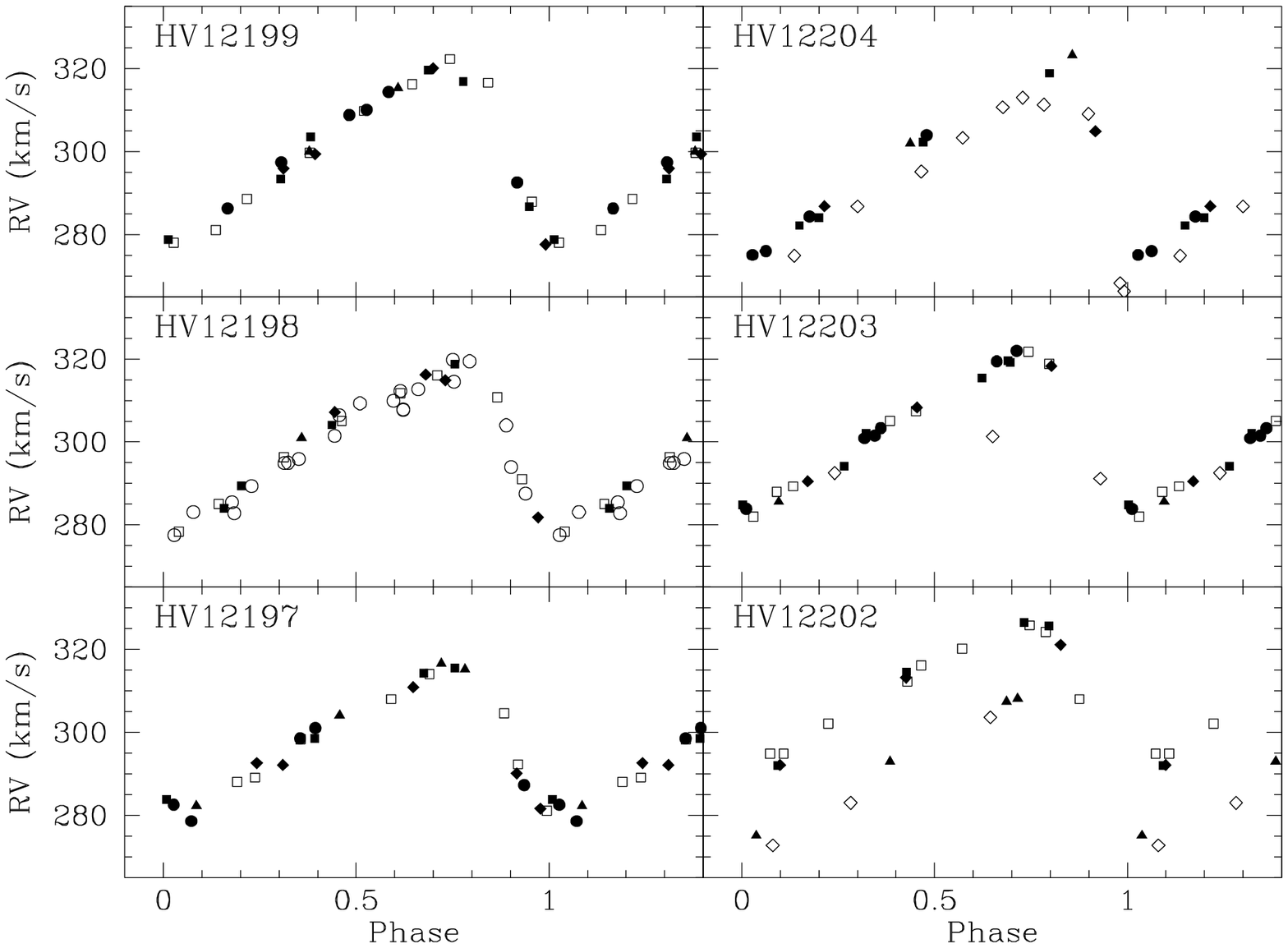}
\figcap{\label{fig.ngc1866_1} The radial velocity curves for six NGC1866
Cepheids. Our new data are marked with filled symbols, while data from
the literature have been over-plotted as open symbols.
The data are marked in the following way: 
LCO 1995: filled squares;
LCO 1996: filled diamonds;
LCO 2000: filled circles;
CTIO: filled triangles;
Storm et al. (\cite{Storm04a}): open circles;
Welch et al. (\cite{Welch91}) HJD $< 2447520$: open squares; 
Welch et al. (\cite{Welch91}) $2447860 < \mbox{HJD} < 2447877$: open
diamonds.
}
\end{figure*}

\begin{figure*}[htp]
\epsfig{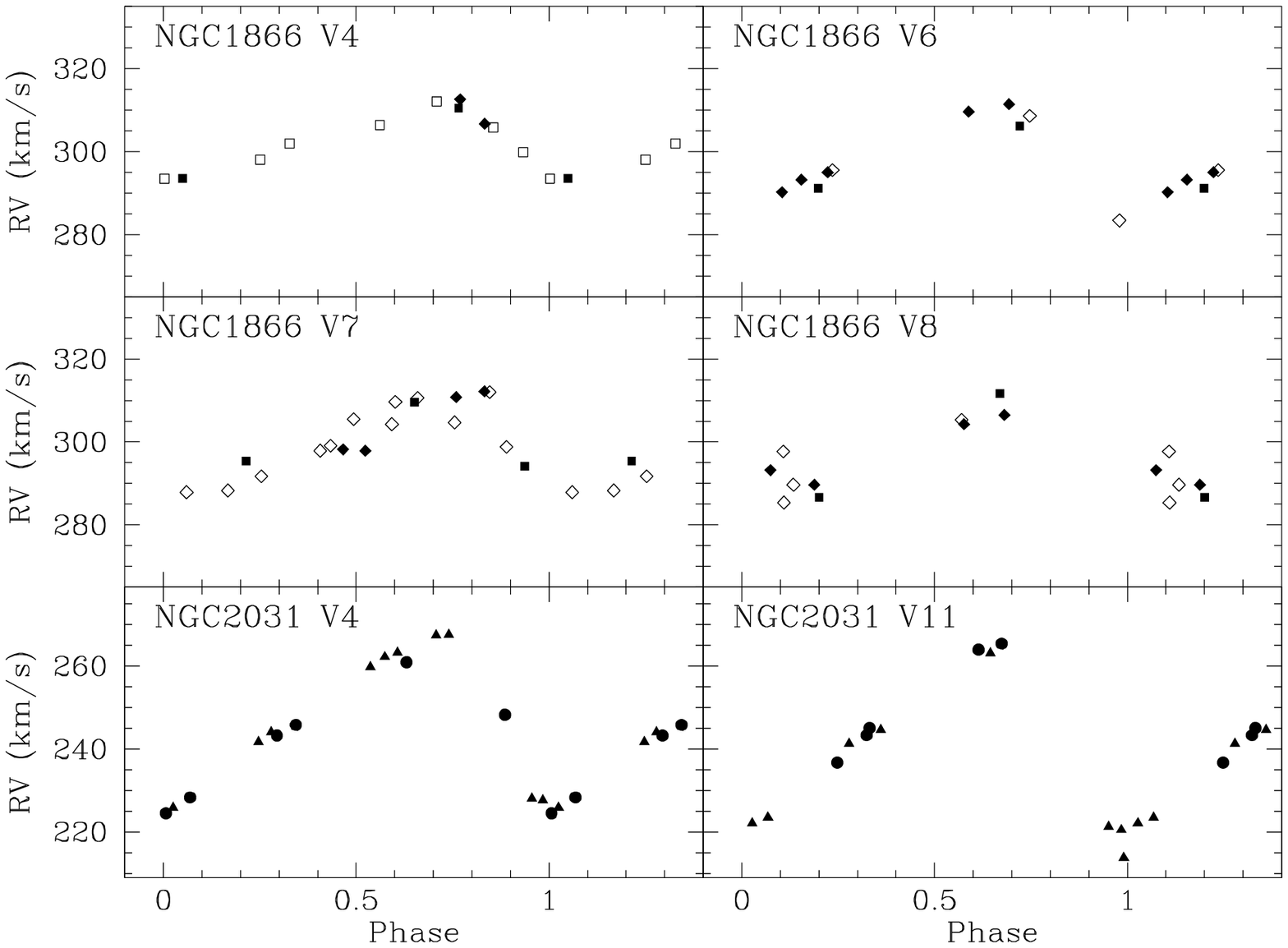}
\figcap{\label{fig.ngc1866_2} The radial velocity curves for four
additional NGC1866 Cepheids and for two NGC2031 Cepheids. The legend is
as in Fig.\ref{fig.ngc1866_1}.}
\end{figure*}

\begin{figure}[htp]
\epsfig{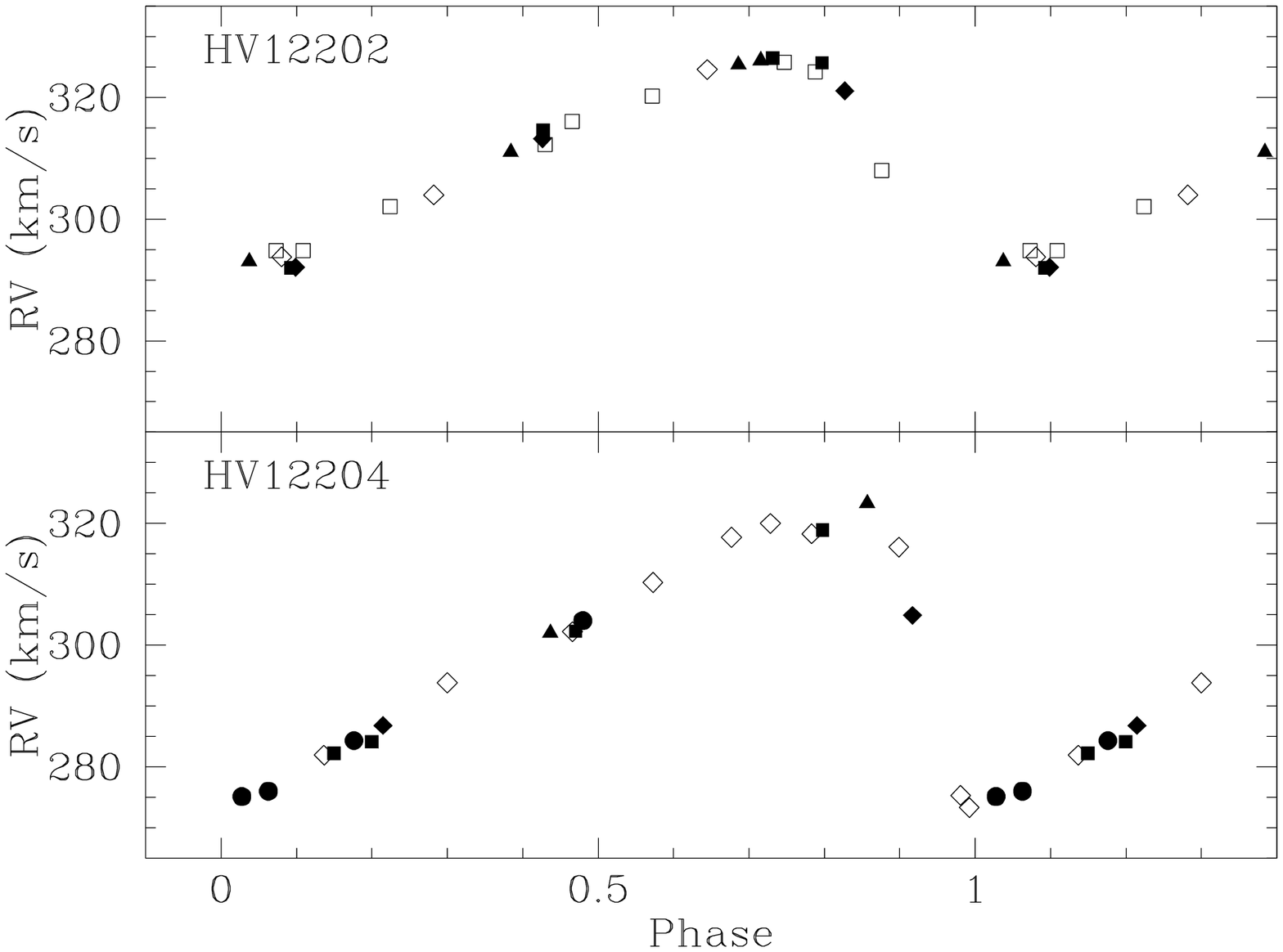}
\figcap{\label{fig.ngc1866shift} The radial velocity curves for the two
NGC1866 Cepheids with suspected orbital motion after the radial
velocities have been shifted as described in the text. The legend is as
in Fig.\ref{fig.ngc1866_1}.}
\end{figure}

  The radial velocity curves are shown in
Figs.\ref{fig.ngc1866_1}-\ref{fig.ngc1866_2}, where our observations are
shown with filled symbols and open symbols are used for data from the
literature. We find excellent agreement with the data from W91
and with Storm et al. (\cite{Storm04a}) for most of the
stars. For HV12203 it appears that one of the observations from W91
is a misidentification (phase=0.68) and it has been rejected in
the following. Two stars, HV12202 and HV12204 show a very systematic shift
between observing runs, and we strongly suspect this is due to orbital
motion. The case for HV12202, as earlier noted by W91, is particularly
strong as the offset is
present in several observing runs. In the case of HV12204 the shift is
only seen for the W91 data. On the other hand
we see no reason why the velocities for a single star should be
systematically off when the rest of the observations from that run are in good
agreement with the other runs. To provide a radial velocity curve which
is useful for a Baade-Wesselink analysis we have determined the shifts
for HV12202 to be $+21$~\kms for the W91 data from the period
$2447860 < \mbox{HJD} < 2447874$ and our observations from the period
$2451571 < \mbox{HJD} < 2451575$ should be shifted by $+18$~\kms.
In the case of HV12204 the W91 data from the same period as for HV12202
should be shifted by $+7$~\kms.
The resulting radial velocity curves are plotted in
Fig.\ref{fig.ngc1866shift}.

  We have determined the systemic velocity, $V_{\gamma}$, for each star
as the phase weighted mean velocity. These values
are tabulated in Tab.\ref{tab.gamma}. For the two stars showing orbital
motion, HV12202 and HV12204, we have performed the phase weighting
on the original, unshifted, data. However, due to the limited sampling
of the orbital motion a significant, but unknown, uncertainty remains
on the systemic velocity for these stars.

  For the two Cepheids in NGC~2031 and for the four NGC1866 stars,
V4, V6, V7, and V8, we have determined rough ephemerides
as tabulated in Tab.\ref{tab.ephem} on basis of the radial velocity
curves. The epochs of maximum light have
been estimated by forcing the minimum radial velocity to occur at 
phase 0.0, which seems reasonable from the radial velocity curves for the
other stars in Fig.\ref{fig.ngc1866_1} and \ref{fig.ngc1866shift}.

\begin{table}
\caption{\label{tab.gamma} The systemic velocity, $V_{\gamma}$, computed
as the phase weighted mean velocity. The random uncertainty is
of the order 0.2~\kms and the systematic error of the order 1~\kms.}
\begin{tabular}{r l}
\hline\hline
\multicolumn{1}{c}{ID} & \multicolumn{1}{c}{$V_{\gamma}$} \\
 & [\kms] \\
\hline
   \object{HV12197} &  298.6\\
   \object{HV12198} &  299.7\\
   \object{HV12199} &  300.4\\
   \object{HV12202} &  $303.1^a$\\
   \object{HV12203} &  302.4\\
   \object{HV12204} &  $296.5^a$\\
       \object{NGC1866-V4} &  302.6\\
       \object{NGC1866-V6} &  298.9\\
       \object{NGC1866-V7} &  298.7\\
       \object{NGC1866-V8} &  298.2\\
       \object{NGC2031-V4} &  247.7\\
       \object{NGC2031-V11} & 242.1\\
\hline
\end{tabular}

{\scriptsize a: Stars with suspected orbital motion.}
\end{table}

\section{The Infrared Surface brightness method}

  The surface brightness method 
was developed in a series of papers by Barnes and Evans
(\cite{BarnesEvans76}) and Barnes et al. (\cite{Barnes76}, \cite{Barnes77}). 
They determined a calibration of the
colour index $(V-R)$ through a linear relation to the
surface brightness parameter $F_V$ defined as
\begin{equation}
\label{eq.Fv}
F_V(\phi) = 4.2207 - V_0 - 0.5\log\theta(\phi)
\end{equation}
\noindent
where $\theta$ is the stellar angular diameter, $V_0$ the de-reddened
visual magnitude, and $\phi$ is the phase.

  The angular diameter is geometrically related to the distance and
stellar radius through
\begin{equation}
\label{eq.theta}
\theta(\phi) = 2R(\phi) / d  = 2 (R_0 + \Delta R(\phi)) / d
\end{equation}
\noindent
where $d$ is the distance and $R$ is the radius. The variation of the
radius can be determined by integrating the radial velocity curve of the
star:
\begin{equation}
\label{eq.deltaR}
\Delta R(\phi)  = \int -p[V_r(\phi)-V_\gamma]d\phi
\end{equation}
\noindent
where $p$ is the so called projection factor converting radial velocity
into pulsational velocity, $V_r(\phi)$ is the radial velocity
and $V_\gamma$ is the systemic velocity.

  As $F_V$ is calibrated from interferometric measurements of stellar
angular diameters it is then possible to solve Eq.\ref{eq.theta} for the
two remaining unknowns, namely the distance, $d$, and the stellar mean
radius, $R_0$. As discussed by Barnes et al. (\cite{Barnes03}) and
S04, it is sometimes necessary to apply a small phase shift, $\Delta \phi$, 
to the radial velocity data to properly phase the spectroscopic and photometric 
data before solving Eq.\ref{eq.theta}.

Welch (\cite{Welch94}) demonstrated the advantages of using the
near-infrared colour index $(V-K)$ together with the $K$-band magnitudes
and made a first calibration of $F_K$ versus $(V-K)$ on the basis of
interferometric diameters of non-pulsating stars. Fouqu\'e and Gieren
(\cite{FG97}) reformulated the near-infrared relation as a $F_V$, $(V-K)$
relation which they calibrated, using an expanded sample of accurate
interferometric measurements of non-pulsating giant stars. Later Nordgren
et al.  (\cite{Nordgren02}) and Kervella et al. (\cite{Kervella04a}) obtained 
very accurate interferometric radii of Cepheids allowing
them to calibrate the surface brightness method directly. For the
Cepheid $\ell$~Car Kervella et al. (\cite{Kervella04b}) made a phase
by phase comparison between their interferometric measurements and
the values returned by the Fouqu\'e and Gieren relation.  Recently
Groenewegen (\cite{Groen04}) has expanded and improved on the work on
the non-pulsating stars. As there is excellent agreement among all these
various calibrations, we choose for consistency with S04 to still adopt
the relationship from Fouqu\'e and Gieren (\cite{FG97}).  In this way
the left hand side of Eq.\ref{eq.theta} seems well established.

  The right hand side of the equation contains the radius variation
which, apart from the direct observables, depends through
Eq.\ref{eq.deltaR} on one non-observable, namely the projection factor $p$.
As an error in $p$ carries directly over in the
derived distance, it is of fundamental importance that $p$ is well
understood. S04 and references therein argue for their choice of a
weakly period dependent $p$, namely $p=1.39 - 0.03 \log P$ where $P$ is
the pulsation period in days. 

\subsection{Results}

  Before applying the ISB method to the observational data we need to
adopt reddenings for the individual stars. The distances derived from
the ISB method are themselves very robust to errors in the assumed
reddening, but the absolute magnitudes of course reflect directly any
errors so introduced.

We have adopted the canonical value of
$E(B-V)=0.06$ which was originally determined by van den Bergh and Hagen
(\cite{vdB68}) from integrated $UBV$ photometry, and by 
Walker (\cite{Walker74}) from stellar $UBV$ photometry.
We assume an uncertainty of about 0.01~mag on this result.

\begin{figure}[htp]
\epsfig{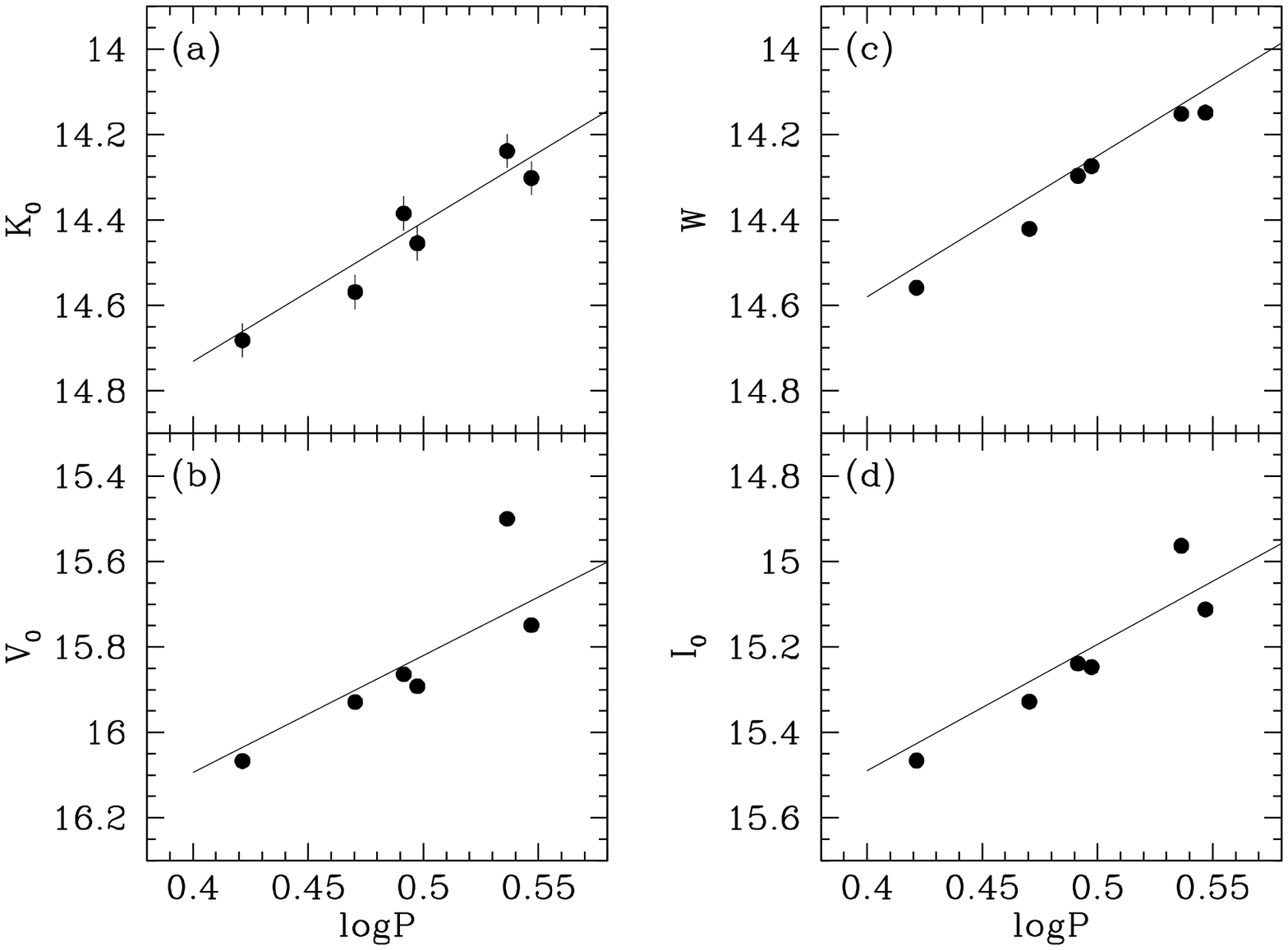}
\figcap{\label{fig.logPVIWK} Panel $a$ shows the observed intensity
averaged $K$ magnitude vs $\log P$ with the Persson et al.
(\cite{Persson04}) relation over-plotted. Panels $b$ to $d$ show the
observed $V$,$I$, and $W$ magnitudes with the corresponding OGLE based
relations from Fouqu\'e et al. (\cite{FSG03}) over-plotted.}
\end{figure}

In Fig.\ref{fig.logPVIWK} we have plotted
the P-L relations in $V$, $I$, $W$, and $K$ using the adopted reddening
of $E(B-V)=0.06$. Over-plotted the data are the observed P-L relations
based on the OGLE data (Udalski et al. \cite{Udalski99}) as determined
by Fouqu\'e et al. (\cite{FSG03}) based on an average LMC reddening of
$E(B-V)=0.10$~mag. The estimated relative random errors on the mean
magnitudes in the optical bands are $0.005$~mag and thus smaller than
the points in the diagram. In the $K$ band where the zero-points are
determined individually for each star the estimated random errors
are somewhat larger, of the order 0.04~mag. 

Groenewegen and Salaris (\cite{Groen03}) have recently claimed a reddening
of $E(B-V)=0.12$ towards NGC1866, but we do not see any support for
this value in our own observational data. From Fig.\ref{fig.logPVIWK}
it is clear that the $V$ and $I$ data appear slightly
too faint (disregarding the slightly peculiar star HV12204) but an
increase of the reddening of only 0.01~mag would make the data in all
the diagrams agree very well with the OGLE based relations. Adopting
an NGC1866 reddening of $E(B-V)=0.12$ on the other hand, would cause
all the data points in the reddening sensitive ($V$ and $I$) diagrams
to appear strongly ($0.15$~mag in $V$) over-luminous with respect to
the OGLE based relations.  We thus argue that the NGC1866 reddening
of $E(B-V)=0.06$ is in good agreement with the reddening scale adopted
by Fouqu\'e et al.  (\cite{FSG03}) which will also be used in the following.
We stress the importance of using self-consistent reddenings
and the superiority of reddening insensitive indices like $W$
and better still the $K$ band.

\begin{table*}
\caption{\label{tab.mM}The measured distance moduli, radii, and absolute
intensity averaged magnitudes for the NGC1866 Cepheids. The estimated
error is just the formal error on the fit. The adopted reddening,
$E(B-V)$, and the phase shift, $\Delta \phi$, applied to the radial
velocity data are also given.}
\scriptsize
\begin{tabular}{r r r r r r r r r r r r r r}
\hline\hline
\multicolumn{1}{c}{ID} & 
$\log P$ & $(m-M)_0$ & $\sigma_{(m-M)}$ & $R$ & $\sigma_R$ & $M_B$ & $M_V$ & $M_I$ & $M_J$ & $M_K$ & $M_W$ & $E(B-V)$ & $\Delta \phi$\\
 & & [mag] & [mag] & [$\Rsolar$] & [$\Rsolar$] & [mag] & [mag] & [mag] &
[mag] & [mag] & [mag] & [mag] & \\
\hline
\object{   HV12199} &   0.421469 & 18.336 & 0.094 &  25.0 &  1.1 & $-1.737$ & $-2.269$ & $-2.870$ & $-3.075$ & $-3.672$ & $-3.777$ &  0.060 & $ 0.025$\\
\object{   HV12203} &   0.470427 & 18.481 & 0.092 &  28.3 &  1.2 & $-1.989$ & $-2.552$ & $-3.153$ & $-3.562$ & $-3.930$ & $-4.060$ &  0.060 & $ 0.050$\\
\object{   HV12202} &   0.491519 & 18.289 & 0.072 &  28.5 &  1.0 & $-1.815$ & $-2.425$ & $-3.050$ & $-3.553$ & $-3.922$ & $-3.992$ &  0.060 & $ 0.025$\\
\object{   HV12197} &   0.497456 & 18.165 & 0.058 &  25.9 &  0.7 & $-1.712$ & $-2.273$ & $-2.918$ & $-3.338$ & $-3.728$ & $-3.891$ &  0.060 & $-0.020$\\
\object{   HV12204} &   0.536402 & 18.202 & 0.044 &  28.3 &  0.6 & $-2.243$ & $-2.702$ & $-3.239$ & $-3.626$ & $-3.981$ & $-4.050$ &  0.060 & $ 0.010$\\
\object{   HV12198} &   0.546887 & 18.314 & 0.028 &  29.8 &  0.4 & $-1.989$ & $-2.565$ & $-3.202$ & $-3.675$ & $-4.030$ & $-4.165$ &  0.060 & $ 0.015$\\
\hline
\end{tabular}
\end{table*}

  On the basis of the observational data presented in the previous
sections we have applied the infrared surface brightness method, using
the Fouqu\'e and Gieren (\cite{FG97}) calibration following the procedure
explained in S04. The derived distance moduli and associated formal
errors for the individual stars have been tabulated in Tab.\ref{tab.mM}
together with the derived absolute magnitudes. We have also tabulated
the phase shifts, $\Delta \phi$, which have been applied to the radial
velocity data to match the photometric angular diameter curve as
discussed by S04.

  For HV12198 we find a distance modulus of $18.31$, which is shorter by
0.11~mag compared to the value of $18.42\pm0.1$ found by Gieren et al.
(\cite{Gieren00b}). This is largely due to the fact that, as discussed
in detail by S04, to avoid effects from possible shocks in the stellar
atmosphere close to minimum radius, we only consider the phase interval
between $0.0$ and $0.8$ here. An additional, but less important, reason is
that we now use the bi-sector for fitting Eq.\ref{eq.theta}.

\begin{figure}[htp]
\epsfig{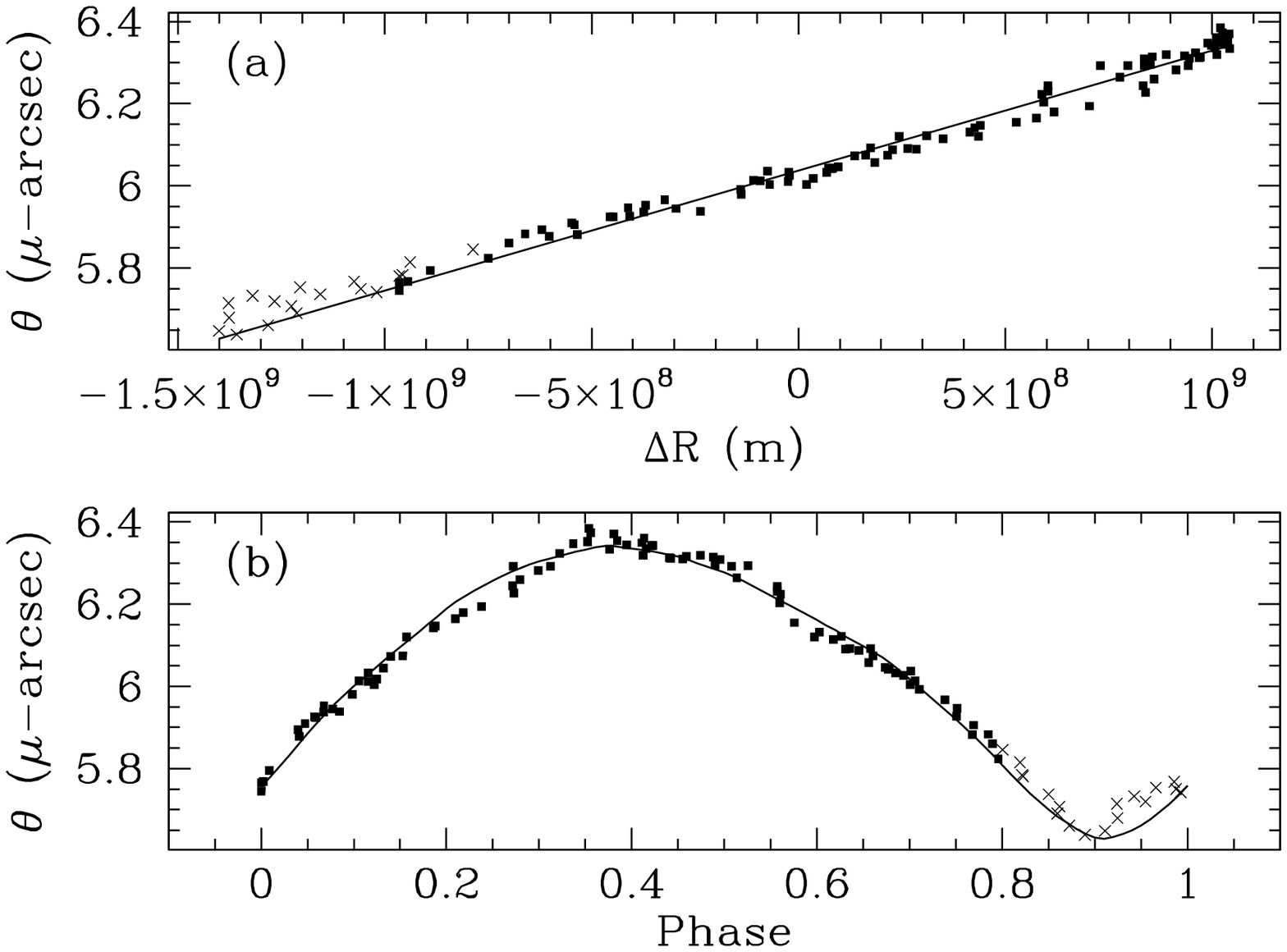}
\figcap{\label{fig.HV12198_fit} Panel $a$ shows the observed data set
of angular diameters, $\theta$ (in micro arc-seconds), versus radius
variation, $\Delta R$, corresponding to Eq.\ref{eq.theta} for the star
HV12198. The straight line is the bi-sector fit to the data in the phase
interval [0.0,0.8]. Panel $b$ shows the angular diameter as a function
of phase. The points correspond to the photometrically determined
angular diameters and the curve corresponds to the integrated pulsational
velocity curve. In both panels the crosses represent discarded points.}
\end{figure}

  In Fig.\ref{fig.HV12198_fit} we show as an example the fit to the data
for HV12198 as well as the angular diameter variation as a function of
phase.

\begin{figure}[htp]
\epsfig{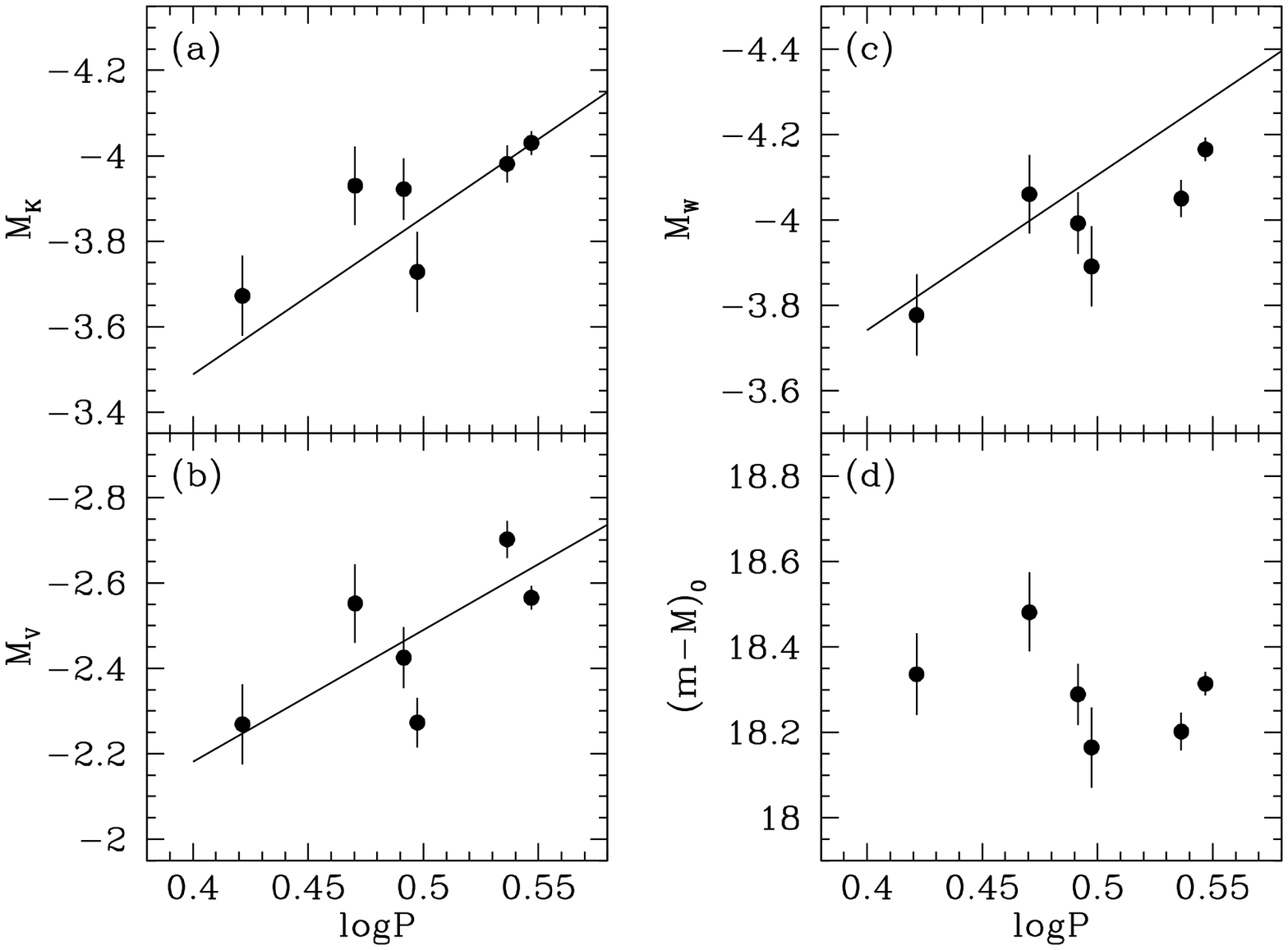}
\figcap{\label{fig.logPVK} Panels $a$ to $c$ show the absolute
magnitudes of the Cepheids versus $\log P$ with the Galactic relation
from S04 over-plotted. Panel $d$ shows the derived distance moduli as a
function of period.}
\end{figure}

In Fig.\ref{fig.logPVK} we have plotted the derived absolute magnitudes in
$V$ and $K$ as well as the Wesenheit index and the distance modulus as a
function of $\log P$ with the Galactic relations over-plotted. It
can be seen that the absolute magnitudes are in good agreement with the
Galactic Cepheids, confirming the finding by S04 from SMC Cepheids that
a metallicity effect on the absolute magnitude of these stars is small,
if at all present. The Galactic relations are largely based on the data
presented in S04 but expanded by four additional stars and some
additional observational data from the literature, see Gieren et al.
(\cite{Gieren05}) for details.
We note that the scatter in Fig.\ref{fig.logPVK} is largely
dominated by errors in the distance estimates as these correlate well
with the deviations seen in the luminosities. The effect of color
(temperature) on the luminosity due to the finite width of the
instability strip is comparatively minor, as can be seen from
Fig.\ref{fig.logPVIWK} which is distance independent and shows a very
low scatter with only HV12204 exhibiting a significant deviation. We
will later argue that this deviation can be explained by a blue
companion.

\section{Discussion}

\subsection{Cluster membership of HV12204}

\begin{figure}[htp]
\epsfig{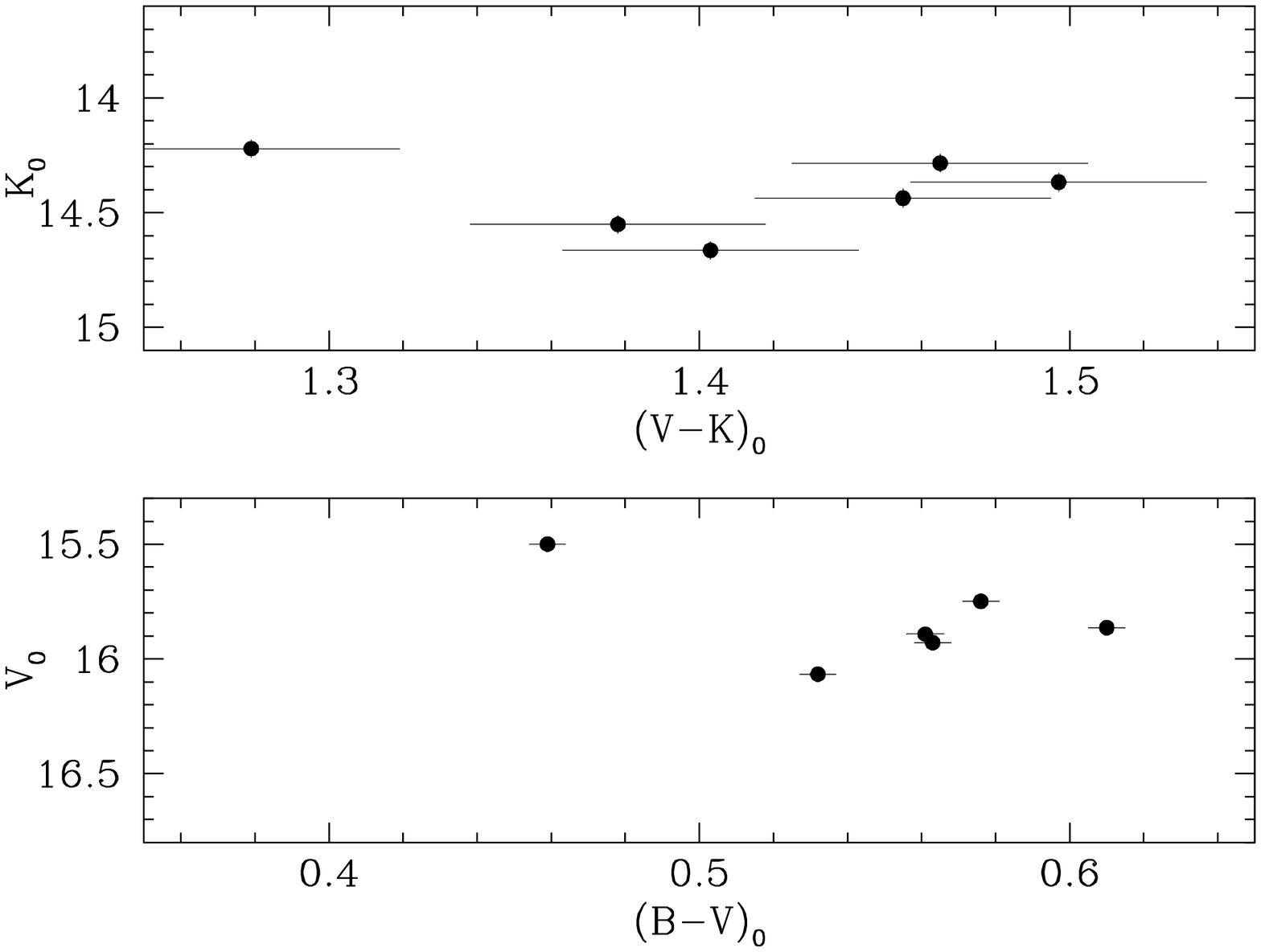}
\figcap{\label{fig.cmd1866} The colour-magnitude diagram for the NGC1866
Cepheids in $V,(B-V)$ and $K,(V-K)$. The bluest object is HV12204.}
\end{figure}

Fischer et al. (\cite{Fischer92}) measured the velocity dispersion of
NGC1866 from non-variable stars to be $2.3$~\kms and found a systemic
radial velocity for the cluster of $V_{\gamma,\mbox{\scriptsize
NGC1866}}=301$~\kms. W91 found a systemic velocity of
$V_{\gamma}=293$~\kms for HV12204 and consequently they suspected it to
be a non-member due to the inconsistent radial velocity and later works
have all dismissed the star on this basis. However, with the additional
radial velocity data we have presented here, it appears that HV12204
is a single-lined spectroscopic binary exhibiting orbital motion. We
do not have enough data to attempt to determine an orbit, but our
data suggest that HV12204 could very well be a member of NGC1866. In
Fig.\ref{fig.logPVIWK} the de-reddened magnitudes of the stars have been
plotted against $\log P$ with the OGLE based P-L relations from Fouqu\'e
et al. (\cite{FSG03}) and from Persson et al. (\cite{Persson04})
over-plotted. It is clear from these plots that the NGC1866 Cepheids
closely follow these relations. In the $V$ and $I$ bands HV12204 appears
slightly over-luminous, but the effect does not seem to be caused by
a difference in distance with respect to the other NGC1866 stars as
the effect changes with wavelength and has disappeared in the $K$
band as well as in the Wesenheit index. Consequently we do not find
sufficient evidence to discard HV12204 as a possible non-member of the
cluster. In Fig.\ref{fig.cmd1866} the colour-magnitude diagram for the
NGC1866 Cepheids has been plotted and it can be seen that HV12204 is the
bluest and brightest of the stars. We cannot exclude that the photometry
is affected by an unresolved blue companion which is further
supported by the fact that the $(J-K)$ color is in good agreement with
that observed for the other cluster Cepheids.

\subsection{The distance to NGC1866 and the estimated distance errors}

  We now have a sample of six NGC1866 Cepheids which we can use in the
following to determine the distance to NGC1866 and to the LMC, and 
to determine the size of the internal random errors of the method.

  The six Cepheids are all within a radius of 3~arcmin from the cluster
center and assuming conservatively a maximum cluster radius of 10~arcmin,
this corresponds to a physical radius of about 150~pc at a distance
of 50~kpc. Thus the maximum error due to depth effects within NGC1866
would be $\pm 150$~pc on a total distance of 50~kpc which is $\pm 0.3$~\%
or $\pm0.007$~mag.  For our purposes we can thus safely assume the stars
to be at the same distance.

  We find a weighted mean distance modulus of $18.28\pm0.05$ with a
standard deviation of 0.11~mag. When we compare the standard deviation
with the formal errors in Tab.\ref{tab.mM}, which range from 0.028 to
0.094, it is obvious that the formal errors significantly (by about
a factor of two) underestimate the true errors. It can also be seen
from the table that the formal errors spread significantly, even though
the data quality is reasonably uniform. In fact the variation of the
estimated errors mostly reflect the overall agreement between the shape
of the two angular diameter curves and to a much lesser extent the random
noise of the data.  In this way a few stars carry a possibly unrealistic
high relative weight. We thus prefer to use the unweighted mean as the
most unbiased estimator, leading to $(m-M)_0 = 18.30\pm0.05$\footnote{This is
our best distance estimate at present, but we caution the reader to take into
account the caveats in subsection \ref{subsec.PLrel}.}, in apparent
excellent agreement with the main-sequence fitting distance
from Walker et al. (\cite{Walker01}).

  Barnes et al. (\cite{Barnes03}) have performed a very thorough
statistical analysis of the optical surface brightness method using
Bayesian techniques. A similar
investigation of the infrared surface brightness method is currently in
press (Barnes et al. \cite{Barnes05}). Preliminary results indeed
suggest that the formal errors of the type reported here are 
underestimates but that the derived individual distances remain unchanged.

\subsection{The LMC distance}

  Even though NGC1866 is some way away from the LMC center we can
attempt, at least formally, to determine the distance to the LMC
on the basis of these data.

  As NGC1866 is a young object it must be close to the LMC disk.
In Fig.\ref{fig.logPVIWK} it seems as if the NGC1866 Cepheids are slightly
(of the order $0.05$~mag) fainter than the average LMC Cepheids in
the $V$ and $I$ bands when disregarding HV12204, but very close to the
mean value in the reddening insensitive $W$ index and in the $K$ band.
As previously discussed the deviation in the $V$ and $I$ bands would
disappear if the reddening were increased by $\Delta E(B-V)=0.01$, well
within the estimated uncertainty of the reddening.  Groenewegen and
Salaris (\cite{Groen03}) have suggested that NGC1866 is more distant than
the surrounding disk population.  At the same time the disk model of  van
der Marel and Cioni (\cite{vandermarel01}) at this position in the north
eastern part of the LMC ({\tt RA:~05:13:39, DEC:~-65:27.9 (2000.0)}) says
that the disk is closer than the LMC center by $0.059$~mag.  These two
effects largely cancel out and confirm our finding that the NGC1866
Cepheids are close to the average LMC Cepheid distance, so we choose
for the present purposes not to apply any correction for geometry,
but add in quadrature an additional uncertainty of 0.05~mag.  We thus
formally find an LMC distance modulus of $(m-M)_{0,\mbox{\scriptsize
LMC}}=18.30\pm0.07$~mag.  We note that the derived modulus is on the
short side of the available estimates,
but as the following discussion will show, this should only be
considered a preliminary estimate of the LMC distance modulus.

\begin{figure}[htp]
\epsfig{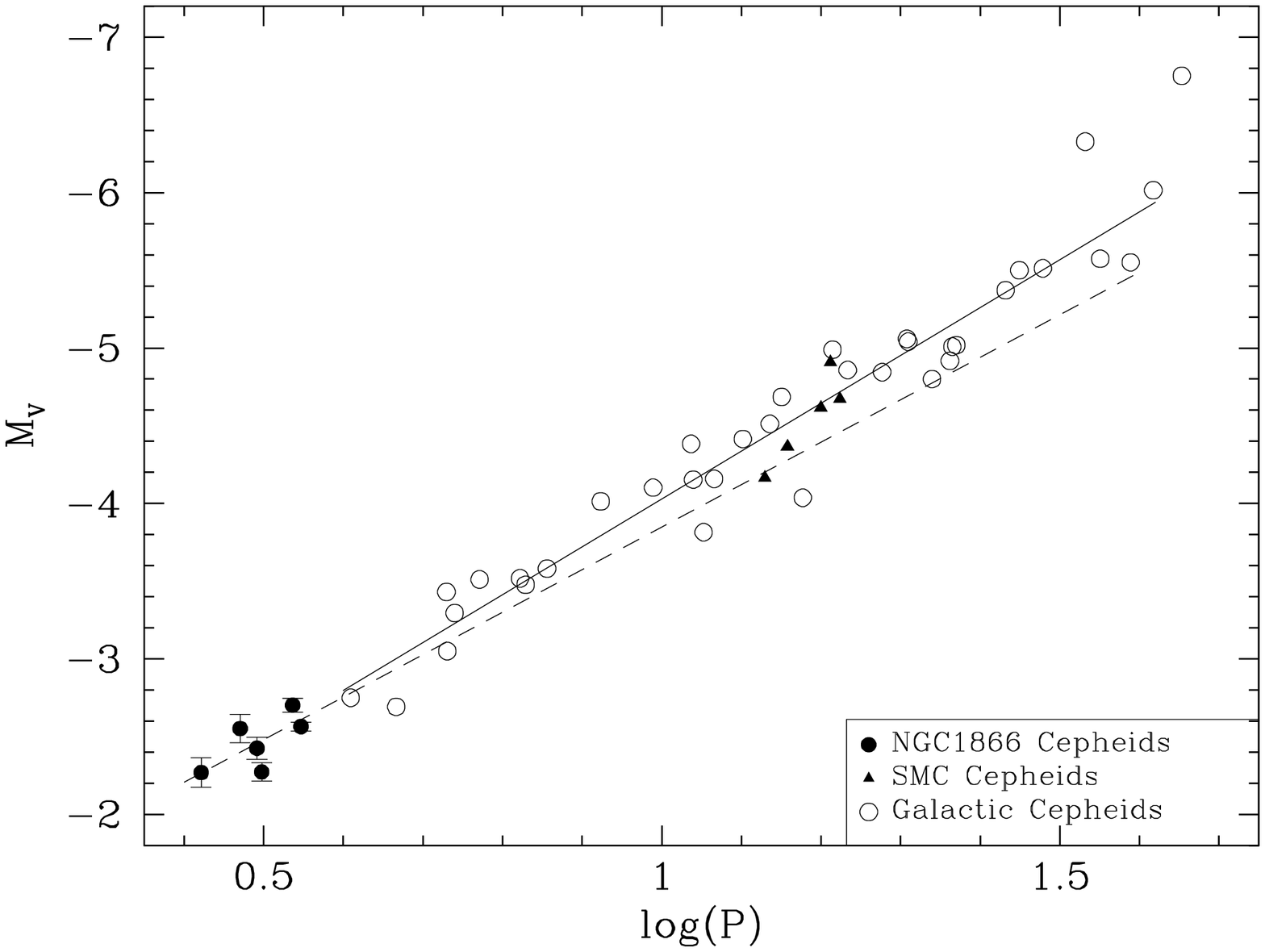}
\figcap{\label{fig.logPMv} The P-L relation for Galactic Cepheids
with the NGC1866 Cepheids overplotted. The
dashed line is the OGLE P-L relation for an adopted LMC distance of $(m-M)_0
= 18.30$~mag.}
\end{figure}

\begin{figure}[htp]
\epsfig{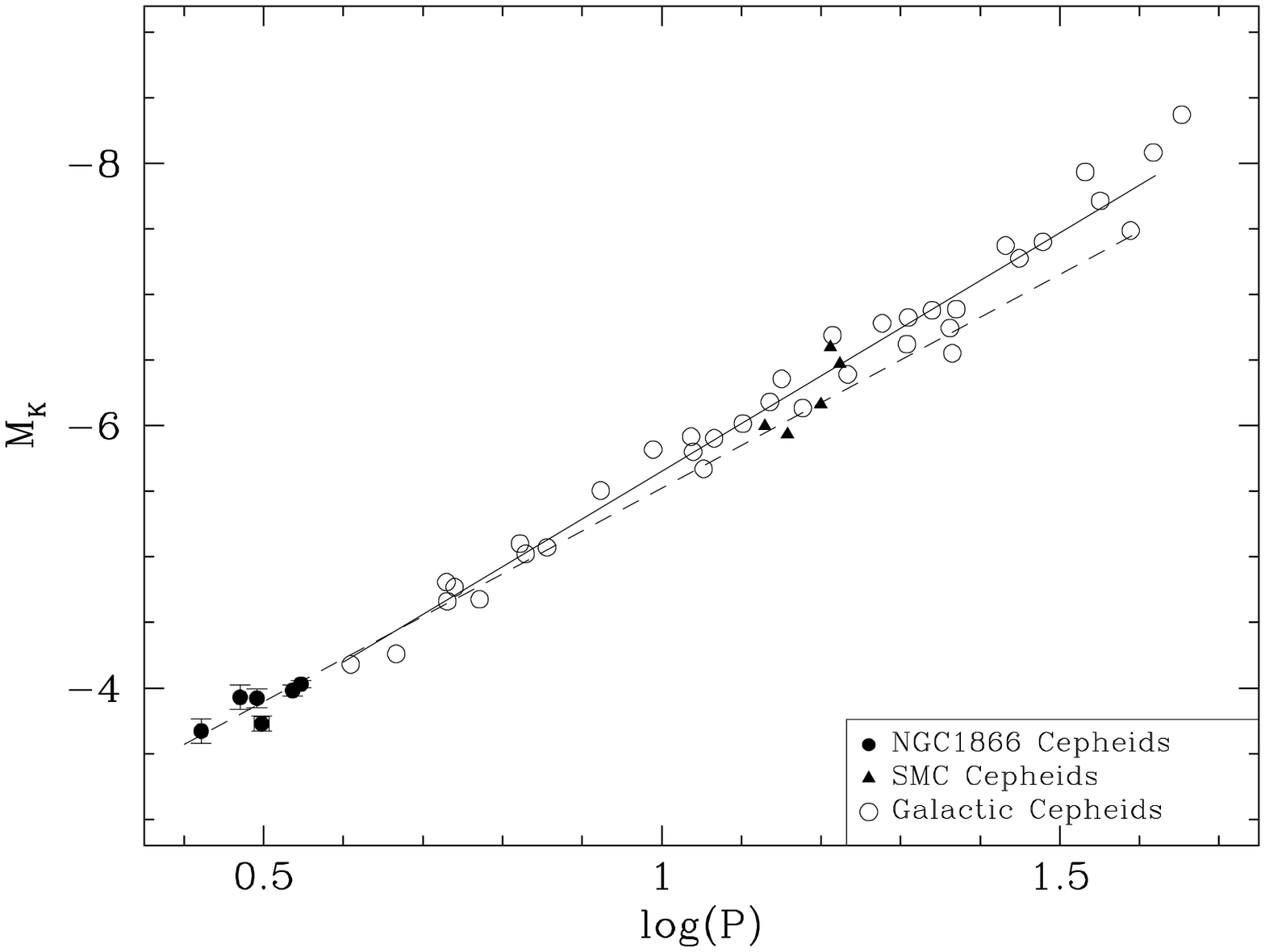}
\figcap{\label{fig.logPMk} The P-L relation for Galactic Cepheids
with the NGC1866 Cepheids overplotted. The dashed line is the
Persson et al. (\cite{Persson04}) P-L relation for an adopted
LMC distance of $(m-M)_0 = 18.30$~mag.}
\end{figure}

\begin{figure}[htp]
\epsfig{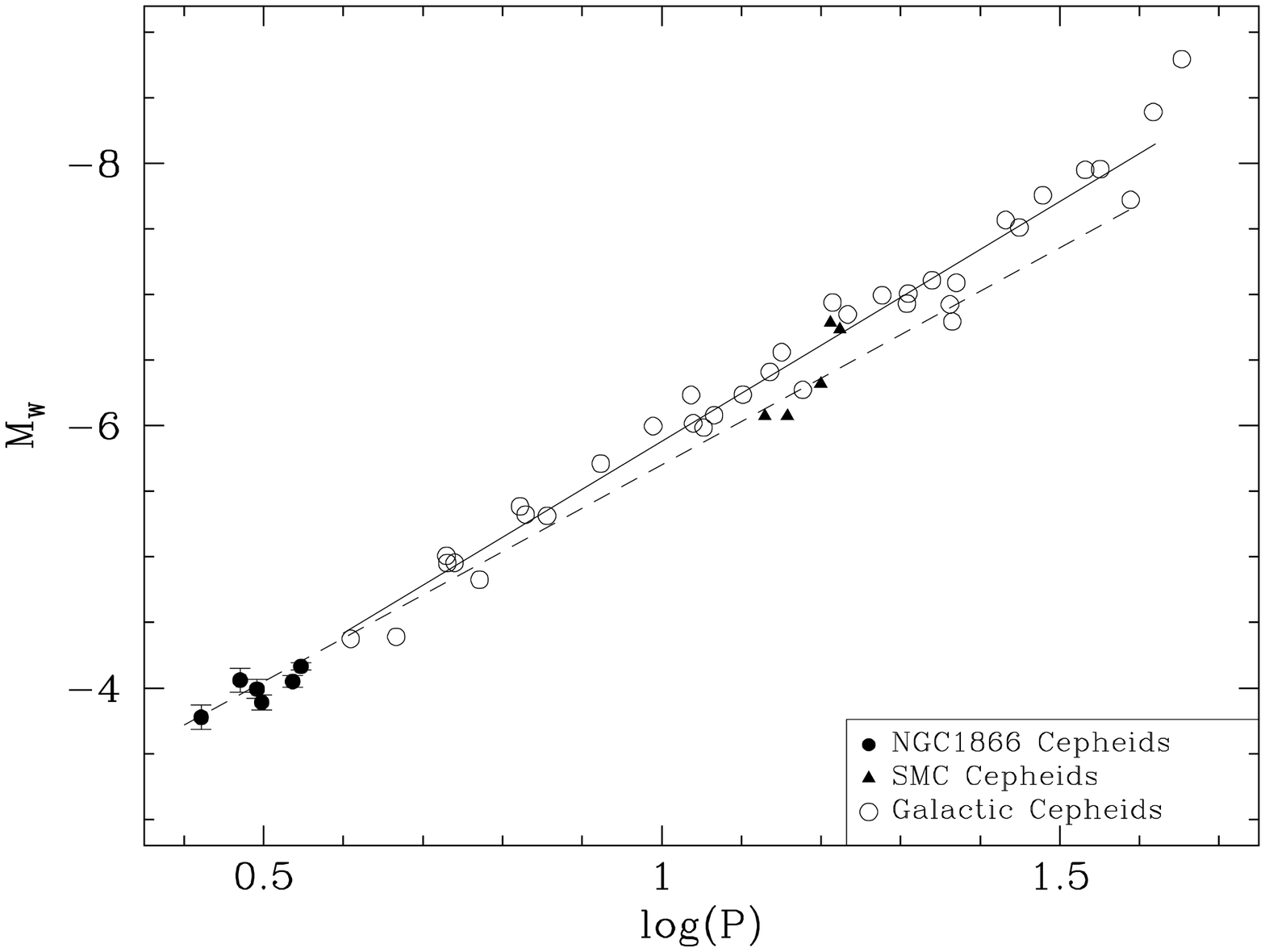}
\figcap{\label{fig.logPW} The P-L relation for Galactic Cepheids
with the NGC1866 Cepheids overplotted. The
dashed line is the OGLE P-L relation for an adopted LMC distance of $(m-M)_0
= 18.30$~mag.}
\end{figure}

\subsection{Comparing the Period-Luminosity relations}
\label{subsec.PLrel}

  The absolute magnitudes which we have derived for the LMC Cepheids
appear in excellent agreement with the values expected from the Galactic
relations derived using exactly the same precepts by S04. In figures
\ref{fig.logPMv}-\ref{fig.logPW} we have overplotted these stars on the
Galactic Cepheids.
We have also overplotted the SMC Cepheids from S04,
which have a metallicity of $\FeH=-0.7$ which is not very different
from the metallicity of the NGC1866 Cepheids of $\FeH=-0.5$ from Hill et
al. (\cite{Hill00}).  The filled lines in these plots represents the linear
Galactic relations fit to the Galactic Cepheids. It is clear that the
low metallicity stars follow this relation quite well and there is no
obvious sign of a difference in slope between the low metallicity and
high metallicity populations. From the perspective of the ISB method
the two populations appear to follow similar P-L relations.

However, the dashed lines overplotted in
Figs.\ref{fig.logPMv}-\ref{fig.logPW} represent the observed apparent P-L
relations shifted to the formal LMC distance of $(m-M)_0 = 18.30$~mag. As
already noted and discussed by Fouqu\'e et al. (\cite{FSG03}), Tammann
et al. (\cite{Tammann03}), and S04 these slopes differ from the slopes
found from the Galactic sample as determined from the ISB method. Now
we can see indications that these slopes also might differ from the
slopes determined directly for the low metallicity stars from the ISB
method. This suggests that there could be a problem either with the
slopes of the apparent P-L relations, which is highly unlikely, or with
the slopes derived from the ISB method. To clarify this point it will
be necessary to study a significantly larger sample of LMC Cepheids
spanning a large range of periods. Gieren et al. (\cite{Gieren05}) will
present the first results from such an expanded sample.

  If we can indeed confirm the suggestion that the LMC and SMC Cepheids
follow the same P-L relation from the ISB method as the Cepheids in
the Galaxy, this will mean that the P-L relation is afterall a
good distance indicator as the slope is not dependent on
metallicity. However, the ISB method itself will then need mending and the
distances and absolute magnitudes derived here and in S04
will need revision.

  It is relevant to note that the theoretical models of
Caputo et al. (\cite{Caputo00}) predict that the metallicity effect on
the luminosity for short period Cepheids is very small, exactly as seen
in Fig.\ref{fig.logPMv}. The predictions for the longer period stars,
and thus for the slope, are tested with the expanded sample of
Gieren et al.  (\cite{Gieren05}).

\section{Conclusions}

  We have presented new near-IR light curves for six NGC1866 Cepheids
and high precision radial velocity curves for six NGC1866 Cepheids and
two Cepheids in NGC2031.

  Using these data we have shown that HV12204 is most likely a member of
NGC1866 but subject to orbital motion, just like HV12202.

  Following the precepts in S04 we have
derived absolute magnitudes and distances to the six NGC1866 Cepheids.
The distances exhibit a standard deviation of 0.11~mag suggesting that the
formal error estimates on the individual distance moduli
are underestimated by about a factor of two.

  Formally we find a distance modulus of $(m-M)_0 = 18.30\pm0.05$ to
NGC1866, but due to the potential problems with the method in its
present form uncovered here, this should be regarded as a preliminary
estimate. Similarly the formal LMC modulus of $(m-M)_0 = 18.30\pm0.07$
might need revision.

  The absolute magnitudes $M_V$, $M_K$ and the Wesenheit index $M_W$ for
the NGC1866 Cepheids are all in excellent agreement with the P-L relations
determined by S04 for Galactic Cepheids. 

  The NGC1866 Cepheids with a metallicity of $\FeH=-0.5$ and the SMC
Cepheids from S04 with a metallicity of $\FeH=-0.7$ appear to
follow largely the same P-L relation as the Galactic Cepheids, in particular they
seem to follow a similar slope.  However, this slope is different from
the slope of the directly observed apparent P-L relations in $K$, $V$,
and $W$ suggesting an intrinsic systematic error with the ISB method
which casts doubt on the distances and absolute magnitudes derived by
the method. On the other hand if this problem originates with the ISB
method itself, this might resolve the very unpleasant problem of the
slope of the P-L relation depending on metallicity.

\section*{Acknowledgements}
  We thank Roeland van der Marel for providing his code for
computing the LMC distance correction for the stars.
This work has benefitted from the use of the McMaster Cepheid database
(http://dogwood.physics.mcmaster.ca/Cepheid/) maintained by D.L.~Welch.
WG gratefully acknowledges financial support from the Chilean Center for
Astrophysics FONDAP 15010003.
OSIRIS is a collaborative project between the Ohio State University and
Cerro Tololo Inter-American Observatory (CTIO) and was developed through
NSF grants AST 90-16112 and AST 92-18449. 

\appendix

\section{Photometric transformations}

\subsection{Transformation to the common instrument system}

  The $JK$ data from all the observing runs were transformed to the Las
Campanas 1998 instrument system before being transformed to the standard
system. 

\subsubsection{The Las Campanas and CTIO data}

  In the case of the Las Campanas observing runs and the CTIO run,
the colour terms for the transformations to the Las Campanas
instrument system were determined using a data set for the LMC cluster
NGC~2136. This field contains a large number of stars without
becoming seriously crowded. The central part containing the cluster
itself was disregarded to reduce the number of stars with systematic
errors due to poor sky subtraction or neighbour subtraction. For each
set of $J$ and $K$ exposures the transformation to the Las Campanas (LCO)
instrument system was determined, assuming relations of the form

\begin{eqnarray}
K_{\mbox{\scriptsize LCO inst}} & = & K_{\mbox{\scriptsize inst}} + 
             \alpha_K (J - K)_{\mbox{\scriptsize inst}} + \beta_K \\
J_{\mbox{\scriptsize LCO inst}} & = & J_{\mbox{\scriptsize inst}} + 
             \alpha_J (J - K)_{\mbox{\scriptsize inst}} + \beta_J 
\end{eqnarray}

  For each observing run the resulting slopes ($\alpha_J$ and
$\alpha_K$) were averaged. The Las Campanas data from the other observing
runs did, as expected, not exhibit significant colour terms, and we have
assumed a value of zero for $\alpha$ in the actual transformations. For
the CTIO data there were slight colour terms of $\alpha_K=-0.03\pm0.01$ and
$\alpha_J = +0.03\pm0.01$.

  The individual $J$ and $K$ data sets were then reanalyzed with the slopes
fixed at the adopted values leaving only the zero point offsets
$\beta_J$ and $\beta_K$ to be fitted. Only stars with estimated errors
less than 0.05~mag were used in this fit and the known variables were of
course excluded. 

\subsubsection{The ESO data}

  NGC2136 was not observed during the ESO observing runs. Instead we
observed 11 standard stars from the list of Carter and Meadows
(1995), during the night Dec.21, 1996, to determine the colour terms.
The stars span a colour range from $(J-K)=-0.02$ to $0.96$.
For the $K$-band we found no
significant colour term and in the $J$-band we found $\alpha_J=+0.06\pm
0.02$. This is in good agreement with the value quoted in the user
manual for this colour range. As the colour range for our science targets
as well as for the comparison stars in these fields are all very similar
($\pm0.2$~mag), the exact value of the colour term is not very important,
as long as it is small. 

  Carter (1993) gives the transformation for normal stars
to the CIT/CTIO system as

\begin{eqnarray}
K_{\mbox{\scriptsize CIT}} & = & K_C - 0.021 (J-K)_C - 0.002\\
(J-K)_{\mbox{\scriptsize CIT}} & = & (J-K)_C - 0.105 (J-K)_C - 0.001
\end{eqnarray}
\noindent
with standard errors of 0.01~mag. Inserting the previously found colour
terms for the ESO data in these equations gives:

\begin{eqnarray}
(J-K)_C & = & (J-K)_{\mbox{\scriptsize ESO}} + 0.06(J-K)_{\mbox{\scriptsize ESO}} \\
        & = & 1.06(J-K)_{\mbox{\scriptsize ESO}} \\
(J-K)_{\mbox{\scriptsize CIT}} & = & (J-K)_C - 0.105 (J-K)_C  - 0.001\\
                              & \approx & 0.895 (J-K)_C \\
                              & = & 0.895 \times
1.06(J-K)_{\mbox{\scriptsize ESO}} \\
                              & = & 0.949 (J-K)_{\mbox{\scriptsize ESO}}
\end{eqnarray}

  This is very similar to the colour equation given by Persson et al.
(\cite{Persson98}) (hereinafter P98)
$(J-K)_{\mbox{\scriptsize CIT}} = 0.954(J-K)_{\mbox{\scriptsize LCO}}$
between the CIT and LCO systems. As the slopes of these
two relations are almost identical and the colour range is very small
($\approx 0.4$~mag) we conclude that we only need to apply a simple offset to
the ESO data to bring the dataset onto the LCO instrument system.

\subsection{Transformation to the CIT standard system}

  On the nights Dec. 27 and 28, 1998 at LCO we observed eight different
standard stars from the list of P98 with a small range in colour
($0.24<(J-K)<0.37$).  The stars were observed several times during
the nights.

  We have performed synthetic aperture
photometry on these stars using the {\tt photcal} package within IRAF and we
determine the zero point shifts $Z_J$ and $Z_K$ to the LCO standard
system (P98) on the basis of the following relations:

\begin{eqnarray}
\label{eq.lcostdJ}
J_{\mbox{\scriptsize LCO}} & = & J_{\mbox{\scriptsize LCO inst}} + k_J X + Z_J \\
\label{eq.lcostdK}
K_{\mbox{\scriptsize LCO}} & = & K_{\mbox{\scriptsize LCO inst}} + k_K X + Z_K
\end{eqnarray}

\noindent

$X$ indicates the airmass. We have adopted the canonical airmass
terms of $k_J=0.10$ and $k_K=0.08$ as quoted by P98 as our data did
not suggest a significantly different value.

  Similarly, synthetic aperture photometry was performed on isolated stars
in the science frames.  Growth curves were fitted for both the science
and standard star data. On the basis of these data
a number of reference stars in each field could be calibrated.
The remaining time series photometry, which was already transformed to the
LCO instrument system as described above, could then be offset
to match these reference stars using the transformations from equations
\ref{eq.lcostdJ}-\ref{eq.lcostdK}.

  The resulting light curves were averaged for each dithered observation set
consisting of typically 5 or 6 exposures. The averaged magnitudes were
finally transformed to the CIT system using the transformation given by
P98:

\begin{eqnarray}
K_{\mbox{\scriptsize CIT}} & = & K_{\mbox{\scriptsize LCO}}\\
(J-K)_{\mbox{\scriptsize CIT}} & = & 0.954(J-K)_{\mbox{\scriptsize LCO}} + 0.015
\end{eqnarray}

  The $K$ and $(J-K)$ photometric measurements are tabulated in
Tab.\ref{tab.JKphot} and the $K$ and $(J-K)$ light curves are plotted
together with the $(V-K)$ colour curve in
Figs.\ref{fig.hv12197vvks}-\ref{fig.hv12204vvks}.

  The $(V-K)$ colour was derived by combining the $V$ data from Gieren et
al. (\cite{Gieren00a}) with the $K$-band data presented here. As the
data were obtained at different times it was necessary to first smooth
the $K$-band data and then interpolate in phase to derive a $(V-K)$
colour corresponding to each $V$ observation. The smoothing was done by
computing the weighted mean of 5 points in phase and magnitude
and interpolating linearly between the resulting points. This approach
results in fewer artifacts than a Fourier fit and thus gives a better
representation of the data. Due to the low amplitude of the $K$-band
light curve any remaining inaccuracies in the smoothed $K$-band light
curve causes only slight changes in the resulting $(V-K)$ curve.

  The photometric errors as returned by DoPHOT are tabulated with the
magnitudes in Tab.\ref{tab.JKphot}. Due to the fact that the stars have
been observed with individual pointings as they did not fit into the
limited field of view of the near-IR cameras, the zero points for each
field have been determined independently using non-variable stars within
the fields. We estimate that the uncertainty on these zero points is
$0.04$~mag.

\section*{References}

\begin{list}{}{\itemindent=-\leftmargin \parsep=0cm}
\raggedright
\bibitem[1967]{AT67}
Arp, H., \& Thackeray, A.D.; \ApJ{1967}{149}{73}
\bibitem[1976]{BarnesEvans76}
Barnes, T.G., \& Evans, D.S.; \MNRAS{1976}{174}{489}
\bibitem[1976]{Barnes76}
Barnes, T.G., Evans, D.S., \& Parsons, S.B.; \MNRAS{1976}{174}{503}
\bibitem[1977]{Barnes77}
Barnes, T.G., Dominy, J. F., Evans, D.S., et al.; \MNRAS{1977}{178}{661}
\bibitem[2003]{Barnes03}
Barnes, T.G., Jefferys, W.H., Berger, J.O., et al.; \ApJ{2003}{592}{539}
\bibitem[2005]{Barnes05}
Barnes, T.G., Storm, J., Jefferys, W.H., Gieren, W.P., \& Fouqu\'e, P.;
\ApJ{2005}{\em In press}{(astro-ph/0506077)}
\bibitem[2002]{Benedict02}
Benedict, G.F., McArthur, B. E., Fredrick, L. W., et al; \AJ{2002}{123}{473}
\bibitem[1993]{Bertelli93}
Bertelli, G., Bressan, A., Chiosi, C., Mateo, M., \& Wood, P.R.;
\ApJ{1993}{412}{160}
\bibitem[1989]{Brocato89}
Brocato, E., Buonanno, R., Castellani, V., \& Walker, A. R.; \ApJS{1989}{71}{25}
\bibitem[2003]{Brocato03}
Brocato, E., Castellani, V., Di Carlo, E., Raimondo, G., \& Walker, A. R.;
\AJ{2003}{125}{3111}
\bibitem[2004]{Brocato04}
Brocato, E., Caputo, F., Castellani, V., Marconi, M., \& Musella, I.;
\AJ{2004}{128}{1597}
\bibitem[2000]{Caputo00}
Caputo, F., Marconi, M., \& Musella, I.; \AandA{2000}{354}{610}
\bibitem[1991]{Cote91}
C\^ot\'e, P., Welch, D.L, Mateo, M., Fischer, P., \& Madore, B.F.; \AJ{1991}{101}{1681}
\bibitem[1997]{FG97}
Fouqu\'e, P., \& Gieren, W.P, \AandA{1997}{320}{799}
\bibitem[2003]{FSG03}
Fouqu\'e, P., Storm, J., \& Gieren, W.P.; 2003, in "Stellar Candles",
Alloin, D., \& Gieren W. (Eds.), Lecture Notes in Physics,
Springer Verlag, 635, 21.
\bibitem[1992]{Fischer92}
Fischer, P., Welch, D.L., C\^ot\'e, P., Mateo, M., \& Madore, B.F.; \AJ{1992}{103}{857}
\bibitem[2001]{Freedman01} 
Freedman, W.L., Madore, B.F., Gibson, B.K., et al.; \ApJ{2001}{553}{47}
\bibitem[2000]{Gibson00}
Gibson, B.K.; \MmSAI{2000}{71}{693}
\bibitem[2000a]{Gieren00a}
Gieren, W.P., G\'omez, M., Storm, J., et al.; \ApJS{2000a}{129}{111}
\bibitem[2000b]{Gieren00b}
Gieren, W.P., Storm, J., Fouqu\'e, P., Mennickent, R., \& G\'omez, M.;
\ApJ{2000b}{533}{L107}
\bibitem[2005]{Gieren05}
Gieren, W.P., Storm, J., Barnes, T.G., et al.;
\ApJ{2005}{\em In press}{(astro-ph/0503637)}
\bibitem[2003]{Groen03}
Groenewegen, M. A. T., \& Salaris, M.; \AandA{2003}{410}{887}
\bibitem[2004]{Groen04}
Groenewegen, M. A. T.; \MNRAS{2004}{353}{903}
\bibitem[2000]{Hill00}
Hill, V., Francois, P., Spite, M., Primas, F., \& Spite, F.;
\AandA{2000}{364}{L19}
\bibitem[1995]{Hilker95}
Hilker, M., Richtler, T., \& Gieren, W.P.; \AandA{1995}{294}{648}
\bibitem[2004a]{Kervella04a}
Kervella, P., Bersier, D., Mourard, D., et al.; \AandA{2004a}{423}{327}
\bibitem[2004b]{Kervella04b}
Kervella, P., Fouqu\'e, P., Storm, J., et al.;
\ApJ{2004b}{604}{L113}
\bibitem[1998]{rvsao98}
Kurtz, M.J., \& Mink, D.J., \PASP{1998}{110}{934}
\bibitem[1992]{Mateo92}
Mateo, M.; \PASP{1992}{104}{824}
\bibitem[2002]{Nordgren02}
Nordgren, T. E., Lane, B. F., Hindsley, R. B., \& Kervella, P.
\ApJ{2002}{123}{3380}
\bibitem[1998]{Persson98}
Persson, S.E., Murphy, D.C., Krzeminski, W., Roth, M., \& Rieke, M.J.; \AJ{1998}{116}{2475}
\bibitem[2004]{Persson04}
Persson, S.E., Madore, B.F., Krzemi\'nski, W., Freedman, W.L.,  Roth, M., \& Murphy, D.C.; \AJ{2004}{128}{2239}
\bibitem[1950]{SN50}
Shapley, H., \& Nail, V.M.; \AJ{1950}{55}{249}
\bibitem[1993]{Schecter93}
Schecter, P.L., Mateo, M., \& Saha, A.; \PASP{1993}{105}{1342}
\bibitem[1988]{Storm88}
Storm, J., Andersen, J., Blecha, A., \& Walker, M. F.; \AandA{1988}{190}{L18}
\bibitem[2004a]{Storm04a}
Storm, J., Carney, B.W., Gieren, W.P., et al.; \AandA{2004a}{415}{521}
\bibitem[2004b]{Storm04b}
Storm, J., Carney, B.W., Gieren, W.P., et al.; \AandA{2004b}{415}{531} (S04)
\bibitem[1968]{vdB68}
van den Bergh, S., \& Hagen, G. L.; \ApJ{1968}{73}{569}
\bibitem[2003]{Tammann03}
Tammann, G.A., Sandage, A., \& Reindl, B.; \AandA{2003}{404}{423}
\bibitem[1979]{Tonry79}
Tonry, J.L. \& Davis, M.; \AJ{1979}{84}{1511}
\bibitem[1999]{Udalski99}
Udalski, A., Soszy\'nski, I., Szyma\'nski, M., et al.;
1999, Acta Astron., 49, 223
\bibitem[2001]{vandermarel01}
van der Marel, R.P., \& Cioni, M.-R.L.; \AJ{2001}{122}{1807}
\bibitem[1987]{Walker87}
Walker, A.R.; \MNRAS{1987}{225}{627}
\bibitem[1974]{Walker74}
Walker, M.F.; \MNRAS{1974}{169}{199}
\bibitem[2001]{Walker01}
Walker, A.R., Raimondo, G., Di Carlo, E., et al.; \ApJ{2001}{560}{L139}
\bibitem[1991]{Welch91}
Welch, D.L., C\^ot\'e, P., Fischer, P., Mateo, M., \& Madore, B.F.;
\AJ{1991}{101}{490} (W91)
\bibitem[1993]{WS93}
Welch, D.L., \& Stetson, P.B.; \AJ{1993}{105}{1813}
\bibitem[1994]{Welch94}
Welch, D.L.; \AJ{1994}{108}{1421}
\end{list}

\end{document}